\documentclass[12pt,preprint]{aastex}

\lefthead{M.~D.~Caballero-Garc\'{\i}a et~al.}
\righthead{GX~339--4}

\received{Date}
\revised{Date}
\accepted{Date}

\journalid{vol}{date}
\articleid{1}{4}
\paperid{id}

\cpright{AAS}{1999}
\ccc{x}

\begin{document}

\title{{\it INTEGRAL} and {\it XMM-Newton} spectroscopy of GX~339--4 during Hard/Soft Intermediate and High/Soft States in the 2007 outburst}

\author{M.~D.~Caballero-Garc\'{\i}a\altaffilmark{1,2},
        J.~M.~Miller\altaffilmark{3},
        M.~D\'{\i}az Trigo\altaffilmark{4},
        E.~Kuulkers\altaffilmark{4},
        A.~C. Fabian\altaffilmark{2},
        J.~M.~Mas-Hesse\altaffilmark{5},
        D. Steeghs\altaffilmark{6,7},
        M.~van der Klis\altaffilmark{8}
        }

\altaffiltext{1}{LAEFF-INTA, P.O. Box 78, 28691 Villanueva de la Ca\~nada, Madrid, Spain}
\altaffiltext{2}{present address: University of Cambridge, Institute of Astronomy, Cambridge CB3 0HA, UK, mcaballe@ast.cam.ac.uk}
\altaffiltext{3}{Department of Astronomy, University of Michigan, 500 Church Street, Ann Arbor, USA, MI 48109, jonmm@umich.edu}
\altaffiltext{4}{ESA/ESAC, Urb. Villafranca del Castillo, P.O. Box 78, 28691 Villanueva de la Ca\~nada, Madrid, Spain}
\altaffiltext{5}{CAB (CSIC-INTA) P.O. Box 50727, 28080 Madrid, Spain}
\altaffiltext{6}{Department of Physics, University of Warwick, Coventry CV4 7AL, UK}
\altaffiltext{7}{Harvard-Smithsonian Center for Astrophysics, 60 Garden Street, Cambridge MA 02138, USA}
\altaffiltext{8}{Astronomical Institute ``Anton Pannekoek'', Kruislaan 403, University of Amsterdam, Amsterdam, 1098 SJ, The Netherlands}


\label{firstpage}

\begin{abstract}

We present simultaneous {\it XMM-Newton} and {\it INTEGRAL} observations of the 
luminous black hole transient and relativistic jet source GX~339--4. GX~339--4 
started an outburst on November of 2006 and our observations were undertaken 
from January to March of 2007. We triggered five {\it INTEGRAL} and three 
{\it XMM-Newton} target of Opportunity observations within this period.
Our data cover different spectral states, namely Hard Intermediate, Soft 
Intermediate and High/Soft. We performed spectral analysis to the data with both 
phenomenological and more physical models and find that a non-thermal component 
seems to be required by the data in all the observations. We find a hardening of 
the spectrum in the third observation coincident with appearance of a
broad and skewed ${\rm Fe}$\,${\rm K}_{\alpha}$ line. In all spectral states 
joint XMM/EPIC-pn,JEM-X, ISGRI and SPI data were fit with the hybrid 
thermal/non-thermal Comptonization model (EQPAIR). While this model accounts 
very well for the high/energy emission observed, it has several drawbacks in the 
description of the lower energy channels. Our results imply evolution in the 
coronal properties, the most important one being the transition from a compact 
corona in the first observation to the disappearance of coronal material in the 
second and re-appearance in the third. This fact, accompanied by the plasma 
ejection events detected in radio on February 4 to 18, suggest that the ejected 
medium is the coronal material responsible for the hard X-ray emission.

\end{abstract}

\section{Introduction} \label{introd}

GX~339--4 (also called V821 Ara) is a recurrent black hole transient X-ray binary 
(i.e. this source undergoes frequent outbursts followed by very faint states) with few episodes
in quiescence (e.g. \cite{kong00}). \citet{hynes03} determined, on the basis of optical spectroscopic 
observations, that this source has an orbital period of 1.7 days and a mass 
function of $f(M)=5.8{\pm}0.5\,M_{\odot}$ (since $f(M){\propto}M\sin^3(i)$, note that this represents 
a lower limit to the black hole mass). The 
distance and the spectral type of the secondary star are uncertain. \citet{hynes04} established the distance to be
6-15\,kpc depending on the spectral type of the secondary (a subgiant), being later than F spectral type.   
Therefore, this black hole transient is a low-mass X-ray binary (LMXB). Slightly later, 
on the basis of binary parameters determination, 
\citet{zdziarski04} suggested the distance to be ${\gtrsim}7$\,kpc, consistent with \citet{hynes04}. 
The inclination of this binary system is known to be low, 
from the small emission-line velocity amplitudes observed. \citet{wu01} determined a plausible value for the
inclination of $i=15^{\circ}$ from spectroscopic data. 
\citet{miller04a}, \citet{miller04b} and \citet{miller06a} found that fits of the relativistic Fe $K_{\alpha}$ line 
imply a very low value of the inclination ($i=20-30^{\circ}$) as well. The latter corresponds to
the inner disk inclination, i.e. not the inclination of the binary plane. 

The timing and spectral properties of GX~339--4 required a
re-interpretation of the standard classification scheme of the black hole states. The behavior of
GX~339--4 supports the view that, as reviewed by \citet{homan01} and \citet{klis06}, X-ray states are important, but are
not a simple function of the luminosity. As claimed by \citet{homan05}, another variable apart from the mass accretion
rate, $\dot{M}$, drives black hole state transitions
and at least two factors must drive state transitions; the second factor may be, e.g., the compactness of the
corona (\citet{homan01}). Black hole X-ray transients and related states (\citet{tanaka95}; \citet{chen97}; \citet{mcclintock06})
provide an excellent opportunity to investigate the importance of irradiation and non-thermal processes in the
X-${\gamma}$-ray spectra of X-ray binaries.
Black hole binary outbursts typically begin and end in the low/hard
state, with a high-energy spectrum that can be roughly described by a powerlaw with a spectral index
${\Gamma}$ varying in the range 1.4-2.1. At some point in the outburst (coinciding with
the presence of radio outflows) the source transits from the hard intermediate (HIMS) to soft intermediate (SIMS) states, characterized
by a quenching of the radio emission (i.e. the so-called jet line). Then, the thermal disk component luminosity increases
and the high energy powerlaw is softer (${\Gamma}=1.5-2.5$) (\citet{miyamoto91a}, \citet{belloni97}, \citet{homanbell05}, see \citet{klis06}
for a review). The transition between these states occurs very rapidly; they were
found in {\it EXOSAT} observations of GX~339--4 by \citet{mendez97}. The Very/High
(hereafter VH) state was first observed in GX~339--4 \citep{miyamoto91a}, when very peculiar X-ray
variability was detected with {\it Ginga/LAC} (6\,Hz QPO, 10-60\,s duration dips and "flip-flops").
\citet{miyamoto91b} even
proposed a jet model in order to account for the observed time variability in this state, on the basis of
the size of the Comptonizing coronae inferred from observations of GX~339--4 in the VH state \citep{miyamoto91a}.
Recent studies (\citet{homan01}, \citet{mendez97}, \citet{wolt07}) point out that the VH state is
really an intermediate state but at a different luminosity level.

Since its diskovery more than 30 years ago by \citet{markert73}, the source has been extensively studied with several
optical, infrared, X-ray and ${\gamma}$-ray observatories. At higher energy its emission has been studied
by all the major observatories, i.e., {\it GRANAT/SIGMA} (\citet{grebenev93}, \citet{bouchet93}), {\it CGRO/OSSE} \citep{grabelsky95},
{\it CGRO/BATSE} (\citet{harmon94}, \citet{rubin98}), {\it ASCA} (\citet{wilms99}),
{\it Ginga} (\citet{ebisawa93} and \citet{ueda94}), {\it RXTE} (e.g. \citet{zdziarski04}, \citet{belloni99}) and {\it INTEGRAL} (\citet{joinet06},
\citet{belloni06}). \citet{homan05} presented a multi-wavelength
study of the 2002 outburst and suggested a non-thermal jet origin for the optical/near-infrared emission
in the low/hard state, while the accretion disk dominates in the high/soft state. This source is very intriguing
in the optical, since its photometric optical behavior shows a very low amplitude modulation superposed
by some flickering events. In addition to the QPOs detected in X-rays (usually appearing in the low/hard state
and maintained in the intermediate states, while quenched in the high/soft state; see e.g. \citet{belloni05}),
this source also shows optical QPOs (${\rm f}{\approx}0.064$\,Hz; \citet{motch82}, \citet{steiman-cameron97}). A very
interesting issue is also the occurrence of optical states corresponding to the X-ray states (\citet{motch83}, \citet{motch85}). 
This kind of variability is difficult to explain if optical radiation comes from reprocessing of X-rays and
seems to demands optical activity originated in the base of a jet.

GX~339--4 may harbor a black hole with a high spin parameter. \citet{miller04a} and \citet{reis08} found 
very small inner disk radius of $2-3\,{\rm R_g}$
(where $R_{g}=GM/c^{2}$ and $a=cJ/GM^{2}$; see \citet{bardeen72} and \citet{thorne74}), implying 
a dimensionless spin parameter of $a>0.8-0.9$ \citep{miller06a}. Relativistic radio jets with $v/c>0.9$ 
have recently been observed, detected \citep{gallo04} during a transition from the low/hard to the high/soft
state, therefore increasing the list 
of known microquasars showing similar radio and X-ray properties. A radio jet is usually reported in the 
low/hard state of GX~339--4, for which \citet{fender99} reported a quenching
of the radio emission during the high/soft state (by a factor of ${\geq}25$ compared to the low/hard state).
The low/hard state would be associated by a steady state of the outflow (\citet{corbel00}, \citet{fender01}, \citet{gallo03}).
It was also proposed that during state transitions the radio emission results in one or more discrete ejection
events, as already observed by \citet{gallo04} during their detection of jets. GX~339--4 shares X-ray 
timing and spectral properties with the classical black hole Cyg~X-1, although the former exhibits more 
frequent state changes and a larger dynamic range of soft X-ray luminosity (\citet{tanaka95},\citet{zdziarski98},
\citet{nowak99}).

While the physical nature of the soft component is commonly 
associated with an optically thick accretion disk, there is no consensus about the origin of the hard 
powerlaw component, where a powerlaw is probably a simplification of a much more complex reality. The 
energy spectra include also additional components, important for the physics of accretion, such as 
the Fe $K_{\alpha}$ fluorescence feature (see, e.g., \citet{miller07}, \citet{reynolds03}) and 
Compton reflection bump, both being different aspects of the same physical 
origin (i.e. fluorescence of the Fe ions and Compton back-scattering, respectively, both being the most obvious 
reactions of an irradiated disk by a high-energy source; \citet{george91}). 
An extremely high ionisation may inhibit the formation of emission and
absorption lines in the inner disk, which would explain the lack of Fe $K_{\alpha}$ fluorescence
emission line in some systems. One important
parameter to measure is the presence/absence of a high energy cut-off in the spectrum, for which
observations at energies $>20$\,keV are necessary. It is known since a long time (\citet{sunyaev80},\citet{grove98}) that 
the spectrum in the low/hard state show a cut-off around $100$\,keV. This cut-off was interpreted
as being a consequence of the thermal (inverse) Comptonization of photons from the disk on a distribution of electrons
thought to represent a corona (for which the geometry is a matter of debate). 

In this paper we report on spectral analysis of simultaneous {\it INTEGRAL} and XMM/EPIC-pn data, during five and
three Target of Opportunity observations of GX~339--4, respectively, occurring between 2007 January 30 and March 31
(epochs 1--5 below; see Table \ref{tobserv} for more details). We fit the obtained spectra with phenomenological models in Section \ref{spec_phenom} and
we observed that the source spectral states evolved from one observation to the other, as reported in Section \ref{state_class}. 
Fits of the spectrum with both phenomenological and Comptonization models are presented in 
Section \ref{spec_phenom}, \ref{spec_eqpair} and \ref{spec_excess} and we end with the discussion of the results in Section \ref{discuss}.

\section{Observations} \label{observ}

In 2006 November, X-ray activity of GX~339--4 was detected with the Rossi X-ray Timing Explorer (RXTE;
\citet{swank06:atel944}). The source had an almost constant flux until end of December 2006, 
when the hard (15-50\,keV) X-ray flux increased by a large amount. It reached its brightest level 
since 2004 November, as detected by SWIFT/BAT \citep{krimm:atel968}.
At the end of January 2007, observations were triggered with the {\it INTEGRAL} satellite \citep{miller:atel980}. 
During the following months, the source underwent an evolution from the low/hard to softer states, as shown in 
preliminary results reported in Caballero-Garcia et~al. 2007a-e.
Observations in radio with ATCA during February 4 to 18 revealed that the source was undergoing a series of plasma 
ejection events \citep{corbel07}. Our {\it INTEGRAL} observations cover 5 Target of Opportunity 
observations of ${\approx}130$\,ks each, spread from 2007 January 30 to March 31 (see Table \ref{tobserv}). 
We also carried out a series of 3 ToO observations with the {\it XMM-Newton} satellite, simultaneous with the second, third and
fifth {\it INTEGRAL} observations. Details about the dates and exposure times appear in Table \ref{tobserv}.

\subsection{INTEGRAL} \label{ins:integral}

The data described were obtained with {\it INTEGRAL},
using the following instruments: the SPectrometer on
{\it INTEGRAL} (SPI; \citet{vedrenne03}), the {\it INTEGRAL} IBIS Soft Gamma-Ray Imager
(ISGRI; \citet{lebrun03}) and the Joint European X-ray Monitor (JEM-X;
\citet{lund03}. SPI, ISGRI and JEM-X allow 
images to be obtained through the coded mask technique.
ISGRI is optimized for 15\,keV to 1\,MeV imaging
and SPI is optimized for high-resolution spectroscopy in the 18\,keV
to 8 MeV band. The former provides an angular resolution of $12'$
full-width half maximum (FWHM) and an energy resolution, $E/{\Delta}E$
of ${\approx}12$ (FWHM) at 100\,keV. SPI provides an angular
resolution of $2.8^{\circ}$ (FWHM) and an $E/{\Delta}E$ of $430$ FWHM
at 1.3 MeV. JEM-X has a fully coded Field of View (FOV) of
$4.8^{\circ}$ diameter and an angular resolution of $3'$ FWHM, with a 
spectral resolution of 1.2\,keV (FWHM) at 10\,keV. JEM-X
has a spectral resolution of $E/{\Delta}E=0.2$ at 6\,keV, thus providing medium
resolution spectral capabilities in the energy range of 3--35\,keV. 

The difference
in the (effective) exposure times between the {\it INTEGRAL} instruments given in Table
\ref{tobserv} (JEM-X, ISGRI and SPI) are due to the difference in the
dead times and variation in the efficiency along the fields of view.
The dithering pattern used during the observations was $5{\times}5$
(square of 25 pointings separated by 2.17 degrees centered on the main
target of the observation); this is the best pattern in order to
minimize background effects for the SPI and ISGRI instruments in
crowded fields. Data reduction was performed using the standard Off-line Science Analysis (OSA) 7.0
software package available from the {\it INTEGRAL} Science Data Centre
(ISDC).  

Because of the steep fall in response off-axis in the case of JEM-X, and because of its
reduced FOV (${\approx}5^{\circ}$ of diameter), we limited the
pointing radius with respect to the GX~339--4 position to be
within $4^{\circ}$. In the case of SPI and ISGRI, with large Fully
Coded Fields of Views (FCFOV) ($16^{\circ}{\times}16^{\circ}$ for SPI
and $9^{\circ}{\times}9^{\circ}$ for ISGRI), pointing selections
were not necessary. In total, 149, 254 and 107 pointings (or Science
Window -- each having exposure times lasting from 1800 to 3600 s)
data were used for SPI, ISGRI and JEM-X, respectively. 
The difference in the number of pointings between SPI and ISGRI is due
to the fact that in epoch 4 and 5 the source was too weak in hard X-rays
($E>20$\,keV) to be detected with SPI. Due to the fact that SPI is a high resolution
spectrograph in its energy range (20-8\,000\,keV), it is not optimized for the detection
of sources without taking into account previous information about
the spatial distribution of them. Thus, we used the list of found 
sources obtained with ISGRI in the (20-40)\,keV mosaic images (see Figure \ref{mosaic})
as our input catalog. This resulted in GX~339--4 as the only source detected
in the 200--300\,keV energy range (Figure \ref{spi_Hchan}) with a detection
significance of 4.8${\sigma}$.

We combined single revolution SPI
spectra from epochs 1--3, considering that there was no significant 
evolution within each revolution (as shown by the relative constancy of the {\it INTEGRAL} and {\it XMM-Newton} light
curves, except for epoch 3 --see below--; Figures \ref{lcurves_xmm} and \ref{lcurves_integral}), 
in order to increase the signal to noise ratio. We obtained 3 SPI spectra 
in total. The same was done 
for the low energy instruments (JEM-X and ISGRI), thus obtaining five 
spectra, namely one for each epoch (see Table \ref{tobserv}). 
We applied 2$\%$ systematic errors to the JEM-X, ISGRI and SPI spectra. We
restricted our analysis in the energy ranges of $5-20$, $18-200$ and
$23-300$\,keV for the JEM-X, ISGRI and SPI spectral analysis, as
recommended\footnote{For OSA software download, {\tt cookbook} documentation and cross-calibration
issues, we refer to the web page: http://isdc.unige.ch/?Soft+download}. 
The SPI and ISGRI spectra were
rebinned at high energies (${\gtrsim}200$\,keV) with the FTOOL
{\tt grppha} procedure to reach the detection level of $2{\sigma}$
per spectral bin. 

\subsection{XMM/EPIC-pn} \label{xmm_ins}

The {\it XMM-Newton} Observatory \citep{jansen01aa} includes three
1\,500~cm$^2$ X-ray telescopes each with an {\it European Photon Imaging
Camera} (EPIC; 0.1--15~keV) at the focus. Two of the EPIC imaging
spectrometers use MOS CCDs \citep{turner01aa} and one uses pn CCDs
\citep{struder01aa}. The {\it Reflection Grating Spectrometer};
\citep[RGS; 0.35--2.5~keV,][]{denherder01aa}
are located behind two of the
telescopes. In addition, there is a co-aligned 30\,cm diameter
Optical/UV Monitor telescope \citep[OM,][]{mason01aa},
providing simultaneous coverage with the X-ray instruments.

GX~339--4 was observed by {\it XMM-Newton} for 16.8~ks on 2007 February 19 
(hereafter called Obs~1 and coinciding with {\it INTEGRAL} Epoch~2), for 17.8~ks on 2007 March 5 
(hereafter called Obs~2 and coinciding with {\it INTEGRAL} Epoch~3) and for 20~ks on 
2007 March 30 (hereafter called Obs~3 and coinciding with {\it INTEGRAL} Epoch~5). We refer
to Table \ref{tobserv} for details.

The thin optical blocking filter was used with the EPIC
camera. The EPIC pn camera was operated in burst mode due to the
high count rate of the source (from $\gtrsim$6700~s$^{-1}$ to
$\gtrsim$3000~s$^{-1}$ for Obs~1 to 3, respectively). For the
same reason, the EPIC MOS cameras were not operated. The individual RGS1 
and RGS2
CCD chips were read out sequentially. This reduces the frame time from 
4.8 to 0.6 seconds
(4.8/8) and consequently the pile-up by almost an order of
magnitude. All the X-ray data products were obtained from the
{\it XMM-Newton} public archive and reduced using the Science Analysis
Software (SAS) version 7.1.0.
In pn burst mode, only one CCD chip (corresponding to a field of
view of 13\farcm6$\times$4\farcm4) is used and the data from that
chip are collapsed into a one-dimensional row (4$\farcm$4) to be
read out at high speed, the second dimension being replaced by
timing information. The duty cycle is only 3$\%$. This allows a time
resolution of 7~$\mu$s, and photon pile-up occurs only for count
rates above 60\,000~s$^{-1}$. Only single and double events
(patterns 0 to 4) were selected. Ancillary response
files were generated using the SAS task {\tt arfgen}. Response
matrices were generated using the SAS task {\tt rmfgen}. 
In this paper we do not use RGS spectra because our main aim
is to make a study of the continuum properties of the spectra.

PN Charge Transfer Inneficiency (CTI) rate dependency has been observed 
for count rates $>200-300$\,cts/s, i.e. 
for very bright sources CTI correction overpredicts the CTI losses. This
can result in an up to 2$\%$ apparent gain shift most visible for spectral
features like edges or lines \footnote{For details see:
http://xmm2.esac.esa.int/docs/documents/CAL-TN-0018.pdf (EPIC
Calibration Status Document), page 4.}. Hereafter,
we applied $2\%$ systematic errors to the EPIC-pn spectra reported 
below. We used EPIC ${\rm pn}$ data in the 0.7--10\,keV energy range.

\subsection{Light curves} \label{lc_phenom}

In Figure \ref{lcurves_swift}, we show the SWIFT/BAT and the RXTE/ASM light curve in the
15--50\,keV and 2--12\,keV energy ranges during the entire outburst.
The horizontal lines shown in the Figure indicate the time intervals
(150 ks each, in five observations) over which our average
spectra were obtained. Note that black hole transients usually
begin and end their outbursts in the low/hard state (see \citet{nowak95};
\citet{fender04}; \citet{homanbell05}), and the 2007 outburst
of GX~339--4 is no exception. As can be seen in Figure \ref{lcurves_swift}, the
beginning of the outburst started earlier at high energies (15--50\,keV)
rather than in softer X-rays (2--12\,keV), when observations of epoch 1
were made. The other four
epochs were made during a period when the source spectrum was much
softer. It is interesting to notice the sudden increase of $15\%$ flux (with respect to the maximum value) 
of the 15--50\,keV light curve, during the
transition from second to third epoch observations,
possibly associated with changes in the high-energy emission source, as we will discuss in Section \ref{discuss}.

In Figure \ref{lcurves_xmm} we show the XMM/EPIC-pn light curves in the 0.6--10\,keV energy range and
calculated hardness ratio (counts in 2--10\,keV band divided by counts in 0.6--2\,keV) with a time binning of
120\,s. We do not see spectral evolution during the time of each observation due to the constancy
of both the light curves and hardness ratio (except for a small increase in the flux during a short time period
in epoch 3, but of the order of ${\leq}6$\%).

In Figure \ref{lcurves_integral} we show the ISGRI (20--200\,keV) light curves during all the observations.
The source was significantly detected during all the epochs with decreasing flux (except for
epoch 3, with higher flux than for epoch 2). Within every epoch the flux was relatively constant, except for
epoch 3, in which the flux was decreasing with time, with a total flux variation of 70\% around
the mean flux value. In spite of this, we combined all the revolution data to get the spectrum
in this epoch as well (as did for the remainder epochs) in order to get higher signal to noise ratio.

\section{Analysis of spectra} \label{spec_analysis}

We performed fits to the combined JEM-X, ISGRI and, when available, SPI 
and XMM/EPIC-pn spectra, for
each of the five epochs (see Table \ref{tobserv}) using XSPEC
\citep{arnaud96} v.11.3. All errors quoted in this work are 90\%
confidence errors, obtained by allowing all variable parameters to
float during the error scan. The hydrogen column density was set to
be free when XMM/EPIC-pn data were available and fixed at 
${\rm N}_{\rm H}=0.5{\times}10^{22}$\,${\rm atom}$\,cm$^{-2}$ 
\citep{dickey90} when only {\it INTEGRAL} data were available.

Our main aim in these fits is to characterize the broad continuum 
thanks to the broad {\it XMM-Newton} and {\it INTEGRAL} spectral coverage. To account for
uncertainties in relative instrument calibrations, we fixed JEM-X
multiplicative calibration constant to 1 and let that of ISGRI and SPI
free to vary in the fits for the different data sets as shown in
Table \ref{param_spec}. In the case of using both INTEGRAL and XMM/EPIC-pn spectra, we
fixed  the EPIC-pn multiplicative constant to 1 and let that of JEM-X, ISGRI 
and SPI free to vary, in order to account for cross-calibration uncertainties
as well \footnote{We restricted the values of the multiplicative constants to 
be $1,{\approx}1{\pm}20\%,1{\pm}30\%,1{\pm}60\%$ for XMM/EPIC-pn,JEM-X, 
ISGRI and SPI relative to XMM/EPIC-pn in order
to obtain reliable fits. When XMM/EPIC-pn data was not available, the same ratio
between the constants was preserved.}. 
The presence of an XMM/EPIC-pn instrumental line was clear in the spectra
of epochs 2 and 3 and we added a gaussian emission line centered at $2.28$\,keV and 
$2.25{\pm}{0.03}$\,keV, respectively. For epoch 5 two absorption instrumental
lines appeared, centered at $1.82$\,keV and $2.14$\,keV, respectively. 
As mentioned in the {\it XMM-Newton} analysis of GRO~J1655--40 \citep{diaz07};
where these features were present as well,
structured residuals near 2.3\,keV are probably due to an incorrect 
instrumental modeling of the Au mirror edges, whilst those near 1.8\,keV are probably due
to an incorrect modeling of the Si absorption in the CCD detectors.


\subsection{Fits of joint XMM/EPIC-pn and INTEGRAL spectra with phenomenological models} \label{spec_phenom}

We performed preliminary fits to the spectra of every single epoch with the phenomenological model
consisting of a multicolor disk black body ({\tt diskbb} in XSPEC) of \citet{mitsuda84} plus a
powerlaw convolved by absorption at low energy in order to describe the high-energy emission. 
In none of the observations this model provided a good fit to the data 
(${\chi}^{2}_{\nu}=46,3.05,4.2,1.8,3.3$ for ${\nu}=96,150,154,46,108$ 
for epochs 1, 2, 3, 4 and 5, respectively). 

In epoch 1 the situation improved dramatically (${\chi}^{2}_{\nu}=2.76$, ${\nu}=94$)
when a high-energy cut-off was included in the model (i.e. {\tt constant*phabs(diskbb + powerlaw)highecut}; ${\rm E}_{\rm c}{\approx}70$\,keV). 
Nevertheless, some positive and negative residuals still remained at 6--10\,keV and 10--20\,keV, 
in the form of a steepening of the spectrum and of an edge, respectively,
indicating the likely presence of a broad ${\rm Fe}$\,${\rm K}_{\alpha}$ line. We included 
a broad ${\rm Fe}$ gaussian line centered at ${\rm E}_{\rm Fe}=6.4-6.97$\,keV, getting the final statistics of 
${\chi}^{2}_{\nu}=1.4$ (${\nu}=91$). With the model {\tt constant*phabs(diskbb+gaussian+powerlaw)highecut},
the final values obtained for the photon index and the folding energy are of ${\Gamma}=1.46^{+0.03}_{-0.02}$ and
$E=66.4^{+0.9}_{-1.1}$\,keV, respectively. These values resemble those found by \citet{belloni06} during the HIMS
of the 2004 outburst.

In epoch 2 low-energy positive residuals in the form of an excess near 1\,keV were 
detected in the XMM/EPIC-pn spectrum and the fit greatly improved when considered (${\chi}^{2}_{\nu}=1.9$, ${\nu}=144$).
The obtained equivalent width is $W_{1\,keV}=11.6{\pm}2.1$\,eV.
This feature is detected in a number of X-ray binaries and has been previously modeled
either as an emission line, or as an edge, and its nature is unclear
(e.g. \citet{kuulkers97}; \citet{sidoli01}; \citet{boirin03}, \citet{diaz07}). However, since this has not been detected
before in GX~339--4 we think that this line probably is an instrumental effect due to incorrect modeling of
${\rm Ne}$\,${\rm IX}$ and ${\rm Ne}$\,${\rm X}$ and ${\rm Fe}$\,${\rm L}$ in the CCD detector. 

In epoch 3 positive and negative residuals appeared, centered at ${\approx}6-10$\,keV and 
${\approx}10-20$\,keV, respectively (see Figure \ref{skewed_Fe}). 
These residuals are compatible with a broad and skewed ${\rm Fe}$\,${\rm K}_{\alpha}$ line and the respective ${\rm Fe}$ edge.
As in epoch 1, we added a gaussian line centered at ${\rm E}_{\rm Fe}=6.4-6.97$\,keV and the fit improved (${\chi}^{2}_{\nu}=3.3$, ${\nu}=146$), but
some residuals in the red wing of the ${\rm Fe}$ line (3--5\,keV) still remained. These residuals point out to the ${\rm Fe}$\,${\rm K}_{\alpha}$ line
having a relativistic origin (i.e. originated very close to the black hole). Thus, we substituted the gaussian component by a {\tt laor} \citep{laor91}
model, which includes general relativity effects in the modeling of the X-ray emission in the inner boundaries of a rotating black hole.
The parameters of this model are the line energy $E$, the disk emissivity index $q$ 
(where the emissivity is assumed to have a power-law form of $J(r){\propto}r^{-q}$; where $q=3$ is expected for a standard disk), the inner
radius of the line emission region in units of $GM/c^2$ (with a lower limit of 1.235\,$R_{g}$ for a maximally rotating black hole), the outer
line emission radius in units of $GM/c^2$ (fixed to the fiducial value of 100\,$R_{g}$), the disk inclination, and the line normalization.
With this model, we obtained significant
better statistics (${\chi}^{2}_{\nu}=2.7$, ${\nu}=143$) and flat residuals in the 3--5\,keV energy range, where the relativistic effects of the ${\rm Fe}$
complex are expected to appear \footnote{Whilst in epoch 1 there are residuals compatible with the broad $Fe$ line, we can not test
if its origin is relativistic, due to the JEM-X lower energy limit of 5\,keV, where the relativistic red wing is expected to appear. Thus, complex models
dealing with the sophistication of the red wing line (as {\tt laor}), can not be applied to the data of epoch 1.}. 
The inner radius of the disk, the emissivity index, the disk inclination and the normalization
were free to vary in the fit. We allowed the value of the inclination to vary in the range 10$^{\circ}$--60$^{\circ}$.
The value obtained for the line energy is very high ($6.83{\pm}0.14$\,keV), indicating a very ionized medium.
The obtained value for the emissivity index is rather steep ($q=5.2^{+0.5}_{-1.0}$), being 
close to the value found in the VH state ($q=4.8-6$ by \citet{miller04a}).
The obtained value for the disk inclination is $i=35^{+4}_{-7}$$^{\circ}$.
High values for the emissivity index indicate that the primary source is very close to the black hole. Steep emissivity indices 
(${\approx}7$) have been observed from a number of Seyfert galaxies (e.g. \citet{fabian04}; \citet{fabian05}). 
The inner radius obtained was rather low ($r_{in}=2.11{\pm}0.11$\,$R_{g}$), thus being 
clearly different from 6\,$R_{g}$ (expected value for a non-rotating Schwarzschild black hole) and close to the value found
by \citet{reis08} ($r_{in}=2-2.1$\,$R_{g}$).
The presence of negative residuals at 10\,keV led us to consider the effects of a smoothed absorption of
the $Fe$ ions in the inner disk. We included a smoothed absorption edge ({\tt smedge} in XSPEC) and the 
fit improved, getting the final statistics of ${\chi}^{2}_{\nu}=1.3$ (${\nu}=141$). Relativistic smearing is expected
to appear in the inner parts of the accreting disk and, with $Fe$ line, accounts phenomenologically for reflection. 
This resulted in an edge energy of $E_{\rm smedge}=8.53{\pm}0.20$\,keV, compatible with 
He-like configuration for $Fe$ and
${\tau}=1.09^{+0.15}_{-0.11}$, with a smeared edge width frozen to $W_{\rm smedge}=5$\,keV (close to the value found by 
\citet{reis08} in the VH state; i.e. $W_{\rm smedge}=4.5^ {+0.5}_{-2.0}$\,keV). 

In epoch 4 some residuals remained in the form of a smeared edge in the 10--20\,keV
energy range. When considered (i.e. model {\tt constant*smedge*phabs(diskbb + powerlaw)}), the fit improved (${\chi}^{2}_{\nu}=1.2$, ${\nu}=44$),
meaning an F-test probability of 1.337E-04 that the improvement was made by chance. The value obtained for the threshold energy  and 
the maximum absorption factor at threshold are $E_{\rm smedge}=9.53^{+0.5}_{-0.6}$\,keV and ${\tau}=1.6{\pm}0.5$, respectively, compatible with
H-like configuration for $Fe$.

Finally, in epoch 5, residuals in the form of an excess at ${\gtrsim}50$\,keV and a smeared edge at
7--10\,keV (${\chi}^{2}_{\nu}=3.3$, ${\nu}=108$) were present. In the residuals there was an emission line like in epoch 2
and we included an emission gaussian line centered at ${\approx}1$\,keV (${\chi}^{2}_{\nu}=2.4$, ${\nu}=105$). The equivalent width
obtained was $W_{1\,keV}=29{\pm}5$\,eV, thus higher than in epoch 2.  
When adding the smoothed
edge (i.e. model {\tt constant*smedge*phabs(diskbb + powerlaw + gaussian)}) the fit greatly improved 
(${\chi}^{2}_{\nu}=1.3$, ${\nu}=104$). The obtained values for both the threshold energy and maximum absorption factor at threshold 
are $E_{\rm smedge}=9.22{\pm}0.25$\,keV and ${\tau}=3.0{\pm}0.5$, respectively, the former compatible with
H-like configuration for $Fe$ \footnote{The edge
for neutral iron is at 7.1 keV, for He-like iron at 8.8 keV, and for
H-like iron at 9.3 keV.}. The value for the absorption factor at threshold is very high and would indicate that it is not due to 
the presence of an absorber, rather being due to reflection effects.
The excess above 50\,keV would indicate an important source of Comptonization by non-thermal particles, as occurring in epoch 1, as shown below. 

The values obtained for the best fit parameters with the best phenomenological
fit are shown in Table \ref{param_spec} and the spectra with the best fit results are shown in Figures
\ref{fit_spec1} and \ref{fit_spec2}. In Table \ref{param_over} we summarize the
values for the photon indices of the powerlaw (analyzed in this Section) and un-absorbed 
luminosities of both the powerlaw and the disk (assuming a distance to the source of 8.5\,kpc).

\subsection{State classification and related behavior}   \label{state_class}

The main features of the spectrum of epoch 1 are a rather flat photon index (${\Gamma}=1.46^{+0.03}_{-0.02}$) and
the clear presence of a high-energy cut-off at $66.4^{+0.9}_{-1.1}$\,keV. The value found for the cut-off is
very close to the value reported by \citet{belloni06} ($E_{c}=70^{+10}_{-8}$\,keV) during HIMS, although the value found
for the photon index in their case was clearly steeper (${\Gamma}=1.92{\pm}0.05$). We conclude that the 
spectrum of epoch 1 corresponds in luminosity (\citet{belloni97}, \citet{fender04}, \citet{homanbell05}) and 
spectral characteristics to the HIMS.
During epochs 2 and 3 the spectra are characterized by
a steep high-energy emission (${\Gamma}{\approx}2.33{\pm}0.02,2.659^{+0.004}_{-0.01}$
for epochs 2 and 3, respectively). In the classification scheme of
\citet{homanbell05}, both correspond to the SIMS.
The value of the photon index from epoch 2 resembles those
reported previously by \citet{nespoli03} (${\Gamma}=2.44{\pm}0.10$) and \citet{belloni06} (${\Gamma}=2.3$)
during SIMS of GX~339--4 as well. The inner disk temperatures found are $0.891{\pm}0.003$\,keV and $0.852^{0.005}_{-0.008}$\,keV,
thus resembling the values found by \citet{malzac06} and \citet{cadollebel05} (i.e. $1.2{\pm}0.2$\,keV) for Cyg~X--1 in Intermediate states as well.
Observations of GX~339--4 during the 2004 outburst by \cite{miller04a} during the VH state
show the same photon index value for the high-energy emission (i.e. ${\Gamma}=2.61^{+0.09}_{-0.01}$) as us for epoch 3, accompanied
by a detection of a relativistic $Fe$ line, as well. Indeed, in the most recent state prescription \citep{homanbell05},
the VH state is the same as the SIMS. 

The spectrum during epochs 4 and 5 is characterized by a 
steep high-energy emission (${\Gamma}=2.39{\pm}0.06,2.65{\pm}0.02$). As can be seen in Figure \ref{lcurves_xmm} 
the hardness ratio (defined as HR=(2-10)\,keV)/(0.2-2)\,keV) of the light curves are very
soft in epochs 4 and 5. The inner disk temperatures found for these epochs are $0.75^{+0.07}_{-0.06}$\,keV and $0.698{\pm}0.003$\,keV, 
respectively, thus resembling the values found in the HS state 
for Cyg X-1 (i.e. 0.6\,keV \citet{gierlinski99}, \citet{frontera01}). The softness, the inner disk temperatures, the values for the photon 
index and the high value for the disk normalization 
component suggest that the source in those epochs is in the
high/soft state, considering the classification scheme of \citet{homanbell05}. The spectral evolution
is presented in Figure \ref{spec_state}. 

As explained above, our observations of GX~339--4 cover the spectral
evolution from the HIMS to high/soft state, passing through the SIMS.
As shown in Section \ref{spec_phenom} and in Table \ref{param_over}, this
spectral evolution involves very significant spectral changes which are strongly correlated. The changes
in the spectral shape are strongly correlated with
the ratio of the powerlaw to the disk luminosity.
In the first epoch there was almost no disk contribution in the spectrum. It showed a cut-off
at ${\rm E}_{\rm c}=66.4^{+0.9}_{-1.1}$\,keV in the powerlaw. The cut-off disappeared in the second epoch
and the powerlaw emission decreased notably, being just $1/7$ of the disk luminosity.
In the third epoch the spectrum hardened again considerably (the powerlaw luminosity being 1.7 times
higher than the disk luminosity, the latter decreased by $50\%$ with respect to the second epoch).
The total luminosity increased by 50\% in the transition from epoch 1 to 2 and remained more or less
unchanged during epochs 2 and 3, while decreasing in epoch 4 and 5 (again decreased by $50\%$).
The luminosities obtained for all the epochs
were within the range of $20-50\%$ the Eddington luminosity
(assuming the compact object being a black hole of $6{\rm M}_{\odot}$).

\section{The high-energy excess} \label{spec_excess}

During the LH state of the GX 339--4 outburst occurring in 1991, a high-energy excess (above 200\,keV) was
observed with OSSE (\citet{johnson93}, \citet{wardzinski02}). In 2004, during the increasing phase
of the 2004 outburst, \citet{joinet06} observed a similar feature with SPI when GX 339--4 was in its
canonical LH state. Similar features have been recently reported for other black hole transients in LH
states observed during the decreasing phase of their outbursts (\citet{kalemci06}, \citet{kalemci05}). It
is also more commonly associated with state transitions \citep{malzac06} and with soft states \citep{gierlinski99}.

Such a hard tail is ussually attributed to the presence of a small fraction of non-thermal
electrons in the hot Comptonizing plasma.
In the HIMS of our observations (epoch 1), GX 339--4 shows a highly significant cut-off
at $66.4^{+0.9}_{-1.1}$\,keV. This indicates that soft photons are Comptonized by a thermal electron
population with a temperature $kT_{e}$ that is related to the cut-off energy. Both for the HIMS (epoch 1) and
SIMS (epochs 2 and 3) non-negligible values of ${\ell}_{nth}/{\ell}_{h}$ were found (as shown below in Section \ref{spec_eqpair}).

In order to investigate further the possible presence of a non-thermal component during the HIMS and SIMS,
IBIS and SPI spectra have been fit with the COMPPS model of \citet{poutanen96} (see Table \ref{fit_compps_param}).
In this thermal Compton model, the spectrum is computed for a homogeneus spherical
hot corona, inside which blackbody photons are injected and then upscattered by hot Maxwellian electrons. The reflected spectrum is
smeared out by rotation of the disk due to special and general relativistic effects using {\tt diskline}-type kernel.
As in the {\tt comptt} model \citep{titarchuk94},
the corona is parametrized by its electron temperature, $T_{e}$, and the radial Thompson optical depth, ${\tau}=n_{e}{\sigma}_{T}H$,
where $n_{e}$ is the electron density, ${\sigma}_{T}$ is the Thompson cross section and $H$ the radius of the sphere.
Using this thermal Comptonization model, some residuals are present at high-energy, especially in the
spectrum of epoch 1. In order to mimic the presence of a non/thermal component, we added a powerlaw
to the pure Comptonization model (see Figure \ref{fit_compps}). This led to an improvement of the ${\chi}^{2}$ that is
significant for epoch 1 (see Table \ref{fit_compps_param}). The F-test probability that this improvement was made
by chance is $p=6.461E-02$ for epoch 1. On the contrary, because of the low statistics
in epochs 2 and 3, adding a powerlaw to the thermal Comptonization model is not required. However, we notice that the
electron temperatures obtained with the thermal Compton plus powerlaw model are rather close to the values obtained with
EQPAIR fits for all the epochs.

\section{Fits of joint XMM/EPIC-pn and INTEGRAL spectra with the hybrid thermal/non-thermal Comptonization EQPAIR model} \label{spec_eqpair}

\subsection{The model}

We saw in the previous Section that a high-energy excess is visible in the spectra. Since this high-energy excess
can not be well described by thermal Comptonization, we go into more sophisticated models containing
a hybrid distribution of thermal/non-thermal particles. From the variety of 
Comptonization models available in the literature, we have chosen the hybrid thermal/non-thermal Comptonization EQPAIR model \citep{coppi99},
and relativistic line emission and disk reflection. The best-fit values for the parameters for all the epochs are shown
in Table \ref{fit_eqpair_param}. Because the presence of an instrumental line at ${\approx}1$\,keV in the XMM/EPIC-pn spectra of epochs 2, 3 and 5
was analyzed in Section \ref{spec_phenom}, this will not be discussed further. 

In the EQPAIR model, emission of the disk/corona system is modeled by a spherical hot plasma cloud with 
continuous acceleration of electrons illuminated by soft photons from the accretion disk. At high-energies
the distribution of electrons is non-thermal, but at low energies a Maxwellian distribution with temperature
$kT_{e}$ is established. 

The disk spectrum incident in the plasma is modeled as coming from a pseudo-Newtonian accretion disk
(${\tt diskpn}$ in XSPEC) extending from $R_{out}=100$\,$R_{g}$ down to the minimum stable orbit
$R_{out}=6$\,$R_{g}$. Its spectral shape is then characterized by the maximum color temperature of the disk,
$kT_{pn}$. Previous observations of Cyg X-1 with X-ray telescopes indicate temperatures ranging from 0.1\,keV
in the LH state up to 0.6\,keV in the HS state (\citet{gierlinski99}, \citet{frontera01}).

The model (EQPAIR) takes into account angle dependence, Compton
scattering (up to multiple orders), photon pair production, pair
annihilation, bremsstrahlung, as well as reflection from a cold disk.
This model includes ionized Compton reflection (as the {\tt pexriv} model in XPSEC; \citet{magdziarz95}).
The reflection component is smeared out by rotation of the disk due to special and general relativistic
effects using {\tt diskline}-type kernel to relativistic motion in the disk 
(as in the {\tt diskline} model in XSPEC; \citet{fabian89}). We used the 
LAOR model \citep{laor91} to model the relativistic iron line emission, with the
emissivity index (${\beta}$) free and tied to the opposite value of that of the EQPAIR.
The disk reflection component is calculated for neutral material with standard
abundances, unless otherwise noted. We asumed free inclination and inner disk radius in the range of $10^{\circ}-60^{\circ}$
and $2-6$\,$R_{g}$. The outer disk radius was fixed to 100\,$R_{g}$.

The EQPAIR model provides the injection of a non-thermal electron distribution
with Lorentz factors between ${\Gamma}_{min}$ and ${\Gamma}_{max}$ and
a powerlaw spectral index ${\Gamma}_{inj}$. 
The system is characterized by the power (i.e. luminosity)
$L_{i}$ supplied by its different components. We express each of them
dimensionlessly as a compactness parameter,
${\ell}_{i}=L_{i}{\sigma}_{\rm T}/(R m_{e}c^{3})$,
where $R$ is the characteristic dimension and ${\sigma}_{\rm T}$
the Thompson cross-section of the plasma. Thus, ${\ell}_{s}$, ${\ell}_{th}$,
${\ell}_{nth}$ and ${\ell}_{h}={\ell}_{th}+{\ell}_{nth}$ correspond to
the power in a soft disk entering the plasma, thermal electron heating,
electron acceleration and the total power supplied to the plasma.
The total number of electrons (not including ${\rm e}^{+}{\rm e}^{-}$ pairs) 
is determined by ${\tau}_{\rm T}$,
the corresponding Thompson optical depth, measured from the center
to the surface of the scattering region. If we consider injection from pairs 
${\rm e}^{+}{\rm e}^{-}$, then the total optical depth of the thermalized 
scattering electrons/pairs is expected to be ${\tau}_{\rm T}{\geq}{\tau}_{\rm P}$.

In general, the spectral shape is insensitive to the exact value of the compactness, but 
it depens strongly on the compactness ratios (${\ell}_{h}/{\ell}_{s}$ and  ${\ell}_{nth}/{\ell}_{h}$).
Thus, we froze ${\ell}_{s}$ to a fiducial value (${\ell}_{s}=10$), as commonly  
reported for other sources with similar characteristics (e.g. \citet{ibragimov05}). In the 
fits reported below, we fit the data with a powerlaw distributed injection of electrons and compare
with the results obtained with a mono-energetic distributed injection of them as well. The former
distribution is expected in the case of shock acceleration of particles, while the second
could be achieved in reconnection events that are expected to power the corona \citep{galeev79}.

\subsection{Fits to the data}

For the first epoch of our observations, due to the low disk contribution, we froze the 
disk temperature to a fiducial value of $kT_{pn}=300$\,eV, as an intermediate value 
between LH and HS values found by \citet{gierlinski99} and \citet{frontera01} for Cyg~X--1. The relativistic 
Fe disk line emission was significant (with a chance probability of 4.823E-05).
The most relevant factor is the inclination
angle of the reflecting material, clearly favouring high values. We obtained worse residuals
and statistics (${\chi}^{2}_{\nu}=1.8$, ${\nu}=82$) for the case of $i=30^{\circ}$ with 
respect to the case of $i=50^{\circ}$ (${\chi}^{2}_{\nu}=1.0$, ${\nu}=82$). 
Since this value is in conflict with the value obtained in Section \ref{spec_phenom} and the
values obtained from fitting the X-ray data in the literature, we will adopt $i=30^{\circ}$ in the fits
of the remainder epochs \footnote{We repeat the same analysis by considering $i=50^{\circ}$ for 
epochs 2, 3, 4 and 5 and the results do not change significantly with respect to the case 
$i=30^{\circ}$.}. High values for the ionization are preferred by the data, with an
improvement of ${\Delta}{\chi}_{\nu}^{2}=0.5$ (for 82 d.o.f) with respect to the neutral case. 
The spectral cut-off already noticed in Section \ref{spec_phenom} can be 
modeled here by Comptonization of thermal distributed particles and reflection from a 
cold disk. As shown in \citet{petrucci01}, with an electron temperature of $kT_{e}=27.5{\pm}1.2$\,keV
and an optical depth of ${\tau}=2.39^{+0.15}_{-0.18}$, it is possible to reproduce the 
spectral cut-off at $66.4^{+0.9}_{-1.1}$\,keV. The spatial distribution of the 
Comptonizing particles is compact (${\ell}_{h}/{\ell}_{s}=3.9^{+0.6}_{-0.3}$) and mildly non-thermal
(${\ell}_{nth}/{\ell}_{h}=0.40^{+0.15}_{-0.03}$). The high-energy distribution is well
fit with a powerlaw injection of non-thermal electrons with relatively low photon 
index (${\Gamma}_{inj}=1.80^{+0.07}_{-0.05}$) with Lorentz factors within the range
1.3--100 (as considered in the remainder epochs as well). This issue means the clear presence of non-thermal
particles in the plasma. When fitting with a mono-energetic distribution of electrons 
with a Lorentz factor of ${\Gamma}=5$, the fit slightly worsened (${\chi}^{2}_{\nu}=1.1$, ${\nu}=83$ 
and worse description of the highest energy emission) with respect to the previous fit 
(i.e. ${\chi}^{2}_{\nu}=1.0$, ${\nu}=82$). 
Albeit not significant, the spectrum seems to be better described by a powerlaw 
injection of non-thermal electrons. Clearly, higher energy data are needed in order to 
disentangle these models.

In the second epoch, XMM/EPIC-pn data allowed us to constrain the disk temperature
($kT_{pn}=897^{+1.4}_{-1.3}$\,eV). The relativistic Fe disk line emission was found to be 
zero (no improvement of the statistics at all, when considered). The data preferred a neutral reflecting disk
(with ${\Delta}{\chi}^{2}=0.15$ of improvement for 142 dof) and the final statistics is
${\chi}^{2}_{\nu}=1.7$ (${\nu}=142$). The statistics is poor and the disk temperature 
very high, which would indicate that EQPAIR model is not adequate enough to describe the softest 
part of the spectrum of this epoch \footnote{As reported in \citet{malzac06} and \citet{cadollebel05},
such high inner temperatures are not realistic for Cyg~X--1. \citet{malzac06} included an additional
warm Comptonization model to the EQPAIR model to describe the soft part of the spectrum
during the Intermediate state finding very good agreement with the data. However, this kind
of work is out of the scope of this paper.}.
Nevertheless, the high energy spectrum is well described
by a purely non-thermal dominated distribution of particles (${\ell}_{nth}/{\ell}_{h}=0.9{\pm}0.1$).
However, the photon index of the powerlaw injection distribution of the Comptonizing electrons
is very steep (${\Gamma}_{inj}=2.87^{+0.05}_{-0.06}$), which indicates the presence of mildly relativistic electrons.
Both the coronal compactness and opacity were found to be very low 
(${\ell}_{h}/{\ell}_{s}=0.05^{+0.003}_{-0.01}$ and ${\tau}_{p}<0.02$) \citep{nowak02}.
Albeit the spectrum does not show significant relativistic smearing, the reflection covering factor was
found to be significant (${\Omega}/2{\pi}=0.40_{-0.04}^{+0.3}$). We obtained the same quality fit when 
considering a mono-energetic distribution of electrons with a Lorentz factor of 
${\Gamma}=5$ (${\chi}^{2}_{\nu}=1.7$, ${\nu}=143$). 

In the third epoch the disk temperature was found to be $kT_{pn}=704{\pm}2$\,eV, thus resembling the 
values found in soft states for Cyg X-1 (\citet{gierlinski99}, \citet{frontera01}). 
Albeit in the phenomenological fits reported in Section \ref{spec_phenom} the presence of a relativistic
iron line was clear in the residuals, we notice the lack of its significance in this model (just a chance probability of 
1.151E-04 compared with 1.040E-09 from the phenomenological fits). This issue can be 
understood by the fact that the {\tt pexriv} model (which is incorporated in EQPAIR to account for reflection) 
is not a proper description of the reflection occurring 
in the inner parts of an accretion disk, since it does not include the Compton scattering (and consequent blurring
in energy) that must occur. The statistics obtained 
is ${\chi}^{2}_{\nu}=1.5$ (${\nu}=138$) and the 
high-energy spectrum is well described taking into account thermal and non-thermal distributed particles 
in the plasma, by the same amount, as occurring in epoch 1. The photon index of the powerlaw distribution
of non-thermal injected particles is very steep (${\Gamma}_{inj}=3.42^{+0.03}_{-0.04}$), meaning a mildly relativistic
population of non-thermal electrons. Both the coronal compactness and the opacity 
(${\ell}_{h}/{\ell}_{s}=0.28^{+0.03}_{-0.01}$ and ${\tau}_{p}=1.43^{+0.03}_{-0.06}$) are as found in \citet{delsanto08}.
The relativistic smearing in this spectrum is important, meaning that reflection probably occurs in the inner 
regions of the accretion disk. High values for the reflection covering factor were favored and we froze this value
(${\Omega}/2{\pi}=1$). 
We considered a mono-energetic injection of electrons case as well (${\Gamma}=5$) and we obtained clearly worse agreement with 
the data (${\chi}^{2}_{\nu}=1.9$, ${\nu}=139$) and with a large discrepancy of the highest energy bins. In this
epoch a powerlaw injection of electrons is clearly favored against the mono-energetic injection scenario.

The results obtained for the fits of epoch 4 resemble those obtained for epoch 3.
The maximum disk temperature was frozen to a fiducial value of
500\,eV, for the same reason as well. When considering a powerlaw injection of non-thermal electrons, the fit statistics 
obtained is ${\chi}^{2}_{\nu}=1.5$ (${\nu}=44$). The corona is relatively compact, mildly non-thermal and relativistic 
smearing is important, thus giving a very high value of the reflection covering factor (${\Omega}/2{\pi}=1$), as
in epoch 3. Again a mono-energetic injection of electrons is a very poor description of the data (${\chi}^{2}_{\nu}=4.5$; ${\nu}=45$)
with a large discrepancy of the highest energy bins. It is worthy to notice as well that in this case the iron line emission
(modeled here as a LAOR component)
was not significant in the data and even worsened the fit when considered (by ${\Delta}{\chi}_{\nu}^{2}=0.3,0.2$ for cases 
of $i=30^{\circ}$ and $i=50^{\circ}$, respectively).

For epoch 5 we could determine the disk temperature ($kT_{pn}=550{\pm}5$\,eV). The relativistic line was significant 
(with a chance probability of 1.029E-06) and the reflection covering factor very
high (${\Omega}/2{\pi}=0.72^{+0.16}_{-0.10}$). The fit statistics was ${\chi}^{2}_{\nu}=1.6$ (${\nu}=99$). As in epochs 1, 
3 and 4, mono-energetic injection of non-thermal electrons is not a good description of the data, due to the worse
statistics (${\chi}^{2}_{\nu}=1.7$, ${\nu}=100$) and discrepancy of the high-energy channels. The observed temperature of the 
disk is cooler (HS state) than in epoch 3 (SIMS state). This fact is accompanied by the increase of the compacticity
for the latter and would indicate that the corona becomes more compact as the disk temperature increases. 

Spectra and models of the five different periods have been plotted in Figure \ref{spec_state}. 

\section{Discussion and conclusions} \label{discuss}

During our observations, we detect spectral evolution from the HIMS to the HS state, passing through SIMS. 
We performed phenomenological and more physical motivated fits to the data and found that during the HIMS
the spectral cut-off at $66.4^{+0.9}_{-1.1}$\,keV is due to the thermal kinetic distribution of electrons in the plasma. 
For the SIMS and HS states we do not find a significant cut-off in the data but applying analysis with 
Comptonization models (EQPAIR and COMPPS) we detect a high-energy tail in all the epochs that can be understood
as evidence of a non-thermal distribution of particles in the plasma. The detection of the non-thermal component
in GX~339--4 in the LH state was done previously by \citet{joinet06} and in the HIMS by \citet{delsanto08}. 
We thus confirm the detection
of the non-thermal component in the SIMS as well (as previously done by \citet{malzac06} for Cyg~X--1).

The results obtained by applying EQPAIR fits to the data indicate a high value for the coronal compactness for epoch 1,
but within the range of values found in the literature. For epochs 3, 4 and 5 this value is high as well (when compared 
to that obtained in epoch 2). We thus confirm the correlation between coronal compactness and covering fraction of the cold
reflecting material by \citet{nowak02} for epochs 2 to 5. The high value of the coronal compactness found for epoch 1 (HIMS)
would indicate that the Comptonizing high-energy source is compact in size. This would be in agreement with the proposed
scenario of \citet{markoff05}, in which the base of the jet could be the source of the Comptonizing elecrons.
The fact that we are detecting the thermal cut-off would be consistent with the detection of the coronal emission as well.
However, for epoch 2 (SIMS), the values for both the coronal compactness and opacity found
are extremely low. Furthermore, the kinetic temperature found for the thermal electrons in the corona 
is very high (and close to the high-energy limit indeed). We understand these as issues indicative of the absence of
coronal emission during this epoch. During epochs 3 to 5 (SIMS to HS), both the coronal compactness and opacity 
increase again (accompanied with the significant detection of a relativistic line in epoch 3), thus indicating re-appearance of
the corona after epoch 2. 
 
The results reported above are supported by the different fits we have done with the hybrid thermal Comptonization
(EQPAIR) model. We fit the data with a powerlaw distributed injection of electrons and compared
with the results obtained with a mono-energetic distributed injection of them as well. The former
distribution is expected in the case of shock acceleration of particles, while the second
could be achieved in reconnection events that are expected to power the corona \citep{galeev79}. We found that the first model 
is better than the second for epochs 1, 3, 4 and 5. This is indicative of a corona of particles distributed at different
speeds being the source of the high-energy emission. However, for epoch 2, albeit the first proposed scenario is not 
discarded, and contrary to the remainder epochs, a mono-energetic distribution of particles is a good description 
of the spectrum. It seems that magnetic reconnection events are driving the high-energy emission in this epoch.

Thus, we conclude that we detect spectral evolution in our data compatible with disappearance of a part \footnote{Since the model still requires a high-energy component.}
of the corona in epoch 2 (SIMS). This was followed by its re-appearance in epoch 3 (SIMS) 
and maintained in epochs 4 and 5. Giving strength to this 
interpretation is the fact that \citet{corbel07} detected a series of plasma ejection events during 4--18 of 
February (2007) in radio, just previously to our observations of epoch 2. Also, the sudden increase in flux in the 15--50\,keV
of the SWIFT/BAT light curve (Figure \ref{lcurves_swift}) of $15\%$ could be related to changes occurring in the source of the 
high-energy emission during the transition from epochs 2 to 3.
The possible disappearance of the corona during epoch 2 resembles what claimed by
\citet{rodriguez08} in the case of GRS~1915+105. This behavior could be understood
as the fact that the ejected medium is the coronal material responsible for the hard X-ray emission.
The spectra of epochs 1, 3, 4 and 5 show a significant fraction of non-thermal particles as well, indicating that
it could be due to other processes apart from thermal Comptonization. For example, synchrotron or self-synchrotron
emission (\citet{markoff03},\citet{markoff05}) occurring at the base of a jet.

\acknowledgments

MCG acknowledges very useful discussion with J. Malzac. This work is based on observations 
made with {\it INTEGRAL}, an ESA science mission with
instruments and science data centre funded by ESA member states and
with the participation of Russia and the USA and on observations with {\it XMM-Newton},
an ESA science mission with instruments and contributions directly funded by ESA
member states and the USA (NASA). We thank to K. Arnaud for providing part of 
the eqpair code used in this work. We also thank E. Jourdain for the valuable discussions 
regarding SPI data analysis and the anonymous referee for the very useful comments, which improved greatly the quality of the paper. 
JMM aknowledges funding from NASA through the INTEGRAL Guest Observer program.
MCG was a MEC funded PhD student during the main work of this paper, supported under grants
PNE2003-04352+ESP2005-07714-C03-03. DS acknowledges a STFC Advanced Fellowship as well as
support through the NASA Guest Observer program. We also acknowledge the Swift/BAT team
for the Swift/BAT transient monitor results.




\clearpage

\begin{deluxetable}{cccccccc}
\tabletypesize{\scriptsize}
\rotate
 \tablecaption{{\it INTEGRAL} and {\it XMM-Newton} Observations LOG and Effective Exposure times of the {\it INTEGRAL} and {\it XMM-Newton} instruments.\label{tobserv}}
 \tablewidth{0pt}
 \tablehead{\colhead{Epoch} & \colhead{XMM-Newton ID} & \colhead{INTEGRAL} & \colhead{XMM-Newton} & \colhead{XMM/EPIC-pn} & \colhead{JEM-X} & \colhead{ISGRI} & \colhead{SPI} 
 }
 \startdata
           &           & (yyyy/mm/dd)  & (UTC hh:mm ; yyyy/mm/dd) &  [s]           &  [s]  &   [s]  &   [s]  \\
           &   &       &            &      &    & & \\
\hline
           &   &       &            &      &    & & \\
         1 &  $-$       & 2007/01/30-02/01 & $-$                & $-$      &  50\,785      &  85\,265      & 98\,324     \\
           &   &       &            &      &    & & \\
         2 & 0410581201 & 2007/02/17-19    & 00:03--04:44 ; 2007/02/19     &  15\,754 & 45\,036       & 73\,356       & 85\,565     \\
           &   &       &            &      &    & & \\
         3 & 0410581301 & 2007/03/04-06    & 11:15--11:15 ; 2007/03/05     &  16\,702 & 49\,187       & 93\,096       & 107\,733    \\
           &   &       &            &      &    & & \\
         4 &   $-$      & 2007/03/16-18    &   $-$              &  $-$     &  50\,966      & 82\,635       & $-$  \\
           &   &       &            &      &    & & \\
         5 & 0410581701 & 2007/03/29-31    & 14:34--20:07 ; 2007/03/30     & 18\,294 &  63\,647      & 93\,655        & $-$  \\
           &   &       &            &      &    & & \\
 \enddata
 \end{deluxetable}


\begin{deluxetable}{cccccc}
\tabletypesize{\scriptsize}
\tablecaption{Parameters obtained for the best fits of the joint XMM/EPIC-pn, JEM-X, ISGRI and SPI spectra (see text for details), using the descriptive 
phenomenological models {\tt constant*phabs(diskbb+gaussian+powerlaw)highecut}, {\tt constant*phabs(diskbb+gaussian+powerlaw)}, 
{\tt constant*phabs*smedge(diskbb+laor+powerlaw)}, {\tt constant*phabs*smedge(diskbb+powerlaw)} and 
{\tt constant*phabs*smedge(diskbb+gaussian+powerlaw)} for the epochs
1 to 5, respectively. See text in Section \ref{spec_phenom} for more details.  \label{param_spec}}
\tablewidth{0pt}
\tablehead{
\colhead{} & \colhead{{\rm Epoch 1}} & \colhead{{\rm Epoch 2}} & \colhead{{\rm Epoch 3}} & \colhead{{\rm Epoch 4}} & \colhead{{\rm Epoch 5}}
}
\startdata
&                                 &           &           &                 & \\
Parameter    &                         &           &           &                &  \\
&                                &           &           &                &  \\
&                                     &  ${\rm Powerlaw}$ &           &                &  \\
&                                  &           &           &                &  \\
${\Gamma}$   &   $1.46^{+0.03}_{-0.02}$       &  $2.33{\pm}0.02$  &   $2.658^{+0.004}_{-0.020}$     &   $2.39{\pm}0.06$             & $2.653{\pm}0.015$  \\
${\rm E}_{\rm cutoff}$   &   $-$       &  $-$  &   $-$     &   $-$     & $-$   \\
$[{\rm keV}]$ &                    &           &            &                &     \\
${\rm E}_{\rm f}$   &   $66.4^{+0.9}_{-1.1}$       &  $-$  &   $-$     &   $-$     & $-$  \\
$[{\rm keV}]$ &                    &           &            &                &     \\
${\rm N}_{\rm pow}$  &  $0.77^{+0.01}_{-0.04}$  &  $0.89^{+0.06}_{-0.03}$  &  $5.99_{-0.3}^{+0.12}$      & $0.82_{-0.16}^{+0.19}$  &  $0.86{\pm}0.05$ \\
$[{\rm ph}/{\rm keV}/{\rm s}/{\rm cm}^{2}]$ at 1 keV &       &                   &           &                  &                         \\
&                                  &           &           &                &  \\
&                                     &  phabs    &           &               &   \\
&                                  &           &           &               &   \\
${\rm N}_{\rm H}$  &  $0.5$ (f) & $0.467{\pm}0.008$ & $0.577{\pm}0.007$ &   $0.5$ (f) &   $0.451{\pm}0.001$\\
$[10^{22}{\rm cm}^{-2}]$ &                           &                                  &                                &                            &                                 \\
&                                  &           &           &                &  \\
&                                     &  diskbb   &           &               &   \\
&                                  &           &           &               &   \\
${\rm kT}_{\rm in}$ &  $0.97{\pm}0.03$            &   $0.891{\pm}0.003$ & $0.852^{+0.005}_{-0.008}$ &  $0.75_{-0.06}^{+0.07}$  &  $0.698{\pm}0.003$ \\
$[{\rm keV}]$ &                             &                     &                           &                   &                    \\
${\rm N}_{\rm bb}$ & $224^{+175}_{-25}$ & $2\,292^{+13}_{-20}$ & $1\,323^{+78}_{-33}$ &  $3\,678{\pm}1\,900$ & $2\,705{\pm}45$    \\
$[({\rm R}_{\rm in}[{\rm km}]/{\rm D}[10{\rm kpc}])^{2}{\times}{\rm cos}{\theta}]$ &                     &                      &                   &                            &                     \\
&                                  &           &           &                &  \\
&                                     &  smedge   &           &               &   \\
&                                  &           &           &               &   \\
$[{\rm E}]$ &       $-$          &  $-$      & $8.53{\pm}0.20$        &  $9.53_{-0.6}^{+0.5}$           & $9.22{\pm}0.25$    \\
$[{\rm keV}]$ &                    &           &            &                &  \\
$[{\tau}]$ &    $-$             &  $-$      &   $1.09^{+0.15}_{-0.11}$      &  $1.6{\pm}0.5$           &  $3.0{\pm}0.5$    \\
&                                  &           &           &               &   \\
&                                     &  ${\rm gaussian}_{\rm 1}\,({\rm emission}\,{\rm ins.})$ &           &               &   \\
&                                  &           &           &                &  \\
${\rm E}_{\rm gauss}$  &  $-$     &  $2.28$ (f)  & $2.25{\pm}0.03$ & $-$ & $-$ \\
$[{\rm keV}]$ &          &              &                 &     &     \\
${\sigma}$  &        $-$          & $0$(f)    &  $0$(f) &    $-$           & $-$   \\
$[{\rm keV}]$ &                     &           &         &                  &       \\
${\rm N}_{\rm gauss}$    &   $-$     & $0.012{\pm}0.007$ & $0.012{\pm}0.003$ & $-$  & $-$ \\
$[{\rm ph}/{\rm cm}^{2}/{\rm s}]$    &           &                   &                   &      &     \\
&                                  &           &           &               &   \\
&                                     &  ${\rm gaussian}_{\rm 2}\,({\rm absorption}\,{\rm ins.})$ &           &               &   \\
&                                  &           &           &                &  \\
${\rm E}_{\rm gauss}$  &   $-$    &  $-$      &   $-$           &    $-$          &  $1.83{\pm}0.02$ \\
$[{\rm keV}]$ &          &           &                 &                 &                  \\
${\sigma}$  &         $-$         &  $-$      &   $-$           &    $-$           & $0$ (f)  \\
$[{\rm keV}]$ &                     &           &                 &                  &         \\
${\rm N}_{\rm gauss}$    &  $-$      &    $-$    &    $-$              &   $-$               & $0.010{\pm}0.004$  \\
$[{\rm ph}/{\rm cm}^{2}/{\rm s}]$    &           &           &                     &                     &                  \\
&                                  &           &           &               &   \\
&                                     &  ${\rm gaussian}_{\rm 3}\,({\rm absorption}\,{\rm ins.})$ &           &               &   \\
&                                  &           &           &                &  \\
${\rm E}_{\rm gauss}$  &  $-$               &   $-$     &   $-$           &   $-$           & $2.15{\pm}0.03$ \\
$[{\rm keV}]$ &                    &           &                 &                 &                 \\
${\sigma}$  &   $-$               &   $-$     &   $-$           &   $-$            &  $0$ (f) \\
$[{\rm keV}]$ &                     &           &                 &                  &                         \\
${\rm N}_{\rm gauss}$    & $-$       &   $-$     &   $-$               &     $-$             & $0.007{\pm}0.002$  \\
$[{\rm ph}/{\rm cm}^{2}/{\rm s}]$    &           &           &                     &                     &                            \\
&                                  &           &           &               &   \\
&                                     &  ${\rm Broad}\,{\rm gaussian}_{\rm 4}\,({\rm Fe})$ &           &               &   \\
&                                  &           &           &                &  \\
${\rm E}_{\rm gauss}$  &  $6.40^{+0.57}$               &   $-$     &        $-$                    &      $-$        & $-$     \\
$[{\rm keV}]$ &                                &           &                               &                 &         \\
${\sigma}$  &   $1.83^{+0.18}_{-0.21}$               &   $-$     &       $-$      &   $-$            &   $-$ \\
$[{\rm keV}]$ &                    &           &                 &                 &                 \\
${\rm N}_{\rm gauss}$     &   $0.042_{-0.01}^{+0.006}$         &   $-$     &   $-$                 &    $-$     & $-$  \\
$[{\rm ph}/{\rm cm}^{2}/{\rm s}]$    &                             &           & $-$                   &            &      \\
&                                  &           &           &               &   \\
&                                     &  ${\rm Laor}$ &           &               &   \\
${\rm E}$  &       $-$                        &  $-$      & $6.83{\pm}0.14$ &  $-$          & $-$\\
$[{\rm keV}]$ &                                &           &                               &                 &         \\
${\it q}$ &      $-$           &  $-$      &  $5.2_{-1.0}^{+0.5}$ &   $-$          & $-$ \\
${\it r_{in}}$ &      $-$           &  $-$      &  $2.11{\pm}0.11$      &   $-$          & $-$ \\
$[{\it R_{g}}]$ &                    &           &                           &                &     \\
${\it r_{out}}$ &      $-$           &  $-$      &  $100$ (f)      &   $-$          & $-$ \\
$[{\rm R_{g}}]$ &                    &           &                           &                &     \\
${\it i}$ &      $-$           &  $-$      &  $35_{-7}^{+4}$      &   $-$          & $-$ \\
$[{\circ}]$ &                    &           &                           &                &     \\
${\rm N}$     &   $-$                               &   $-$     &    $0.031^{+0.005}_{-0.003}$    &    $-$     & $-$  \\
$[{\rm ph}/{\rm cm}^{2}/{\rm s}]$    &                             &           &                       &            &      \\
&                                  &           &           &               &   \\
&                                     &  ${\rm Broad}\,{\rm gaussian}_{\rm 5}$\,(${\approx}$\,1keV) &           &               &   \\
&                                  &           &           &                &  \\
${\rm E}_{\rm gauss}$  &  $-$                           &   $1.18{\pm}0.02$     &       $-$              &   $-$        & $1.13{\pm}0.03$     \\
$[{\rm keV}]$ &                                &           &                               &                 &         \\
${\sigma}$  &   $-$                           &   $0.09{\pm}0.01$     &       $-$               &   $-$            &   $0.185{\pm}0.015$ \\
$[{\rm keV}]$ &                    &           &                 &                 &                 \\
${\rm N}_{\rm gauss}$     &   $-$                               &   $0.07{\pm}0.01$     &   $-$               &    $-$     & $0.106{\pm}0.018$  \\
$[{\rm ph}/{\rm cm}^{2}/{\rm s}]$    &                             &           &                       &            &      \\
&                                  &           &           &               &   \\
&                                  &           &           &               &   \\
&                                  &           &           &               &   \\
&                                     &  Intrumental normalization factors &           &                &   \\
&                                     &           &           &                 &  \\
${\rm C}_{\rm XMM/EPIC-pn}$  &   $-$      &  1.0 (f)  &  1.0 (f)  &     $-$       & 1.0 (f)  \\
${\rm C}_{\rm JEM-X}$  &   1.0 (f)        &   $0.89{\pm}0.01$        & $0.827{\pm}0.009$  &   1.0 (f)     & $0.86{\pm}0.02$ \\
${\rm C}_{\rm ISGRI}$  &  $1.37{\pm}0.02$     & $1.39{\pm}0.02$  & $1.18{\pm}0.02$               &   $1.28{\pm}0.02$  &  $1.38_{\pm}0.02$ \\
${\rm C}_{\rm SPI}$  &  $1.57{\pm}0.03$   &  $1.56{\pm}0.04$         & $1.52^{+0.06}_{-0.07}$ &   $-$          & $-$  \\
&                                     &           &           &                &   \\
 &                                     &           &           &                &   \\
${\chi}_{\nu}^{2}$ &      $1.4$       &  $1.9$   &  $1.3$   &  $1.20$        & $1.3$  \\
${\nu}$ &             $91$             & $144$     &  $141$    &  $44$          & $102$  \\
&                                    &           &           &                &   \\
\hline
\enddata
\tablecomments{Parameters fixed in the fits are denoted by 'f'. EPIC-pn instrumental absorption features are implemented in XSPEC with negative gaussian profiles.}

\end{deluxetable}

\begin{deluxetable}{ccccccc}
\rotate
 \tablecaption{Luminosities (in the 0.7--300\,keV energy range) and parameters (with ${\Gamma}$ and ${\rm E}_{c}$ being the photon index and the cut-off energy,
respectively, see the Section \ref{spec_phenom} for more details).\label{param_over}}
 \tablewidth{0pt}
 \tablehead{\colhead{Epoch} & \colhead{${\rm L}_{disk}$} & \colhead{${\rm L}_{pow}$} & \colhead{${\rm L}_{total}$} & \colhead{${\Gamma}$} & \colhead{${\rm E}_{c}$} & ${\rm F}_{disk}/{\rm F}_{pow}$
 }
 \startdata
           &  [erg/s]   &  [erg/s]       &  [erg/s]  &        & [keV]  &     \\
           &   &       &            &      &     \\
\hline
           &   &       &            &      &     \\
         1 & $5.4{\pm}1.2{\times}10^{37}$ & $2.0{\pm}0.4{\times}10^{38}$  & $2.6{\pm}0.6{\times}10^{38}$ (0.33${\rm L}_{\rm Edd}$) & $1.46^{+0.03}_{-0.02}$ & $66.4^{+0.9}_{-1.1}$    & $0.27$ \\
         2 &  $3.1{\pm}0.7{\times}10^{38}$  &  $4.7{\pm}1.1{\times}10^{37}$    &  $3.4{\pm}0.8{\times}10^{38}$ (0.46${\rm L}_{\rm Edd}$) &  $2.33{\pm}0.02$ &   $-$ & $7.0$ \\
         3 &  $1.3{\pm}0.4{\times}10^{38}$  &  $2.2{\pm}0.6{\times}10^{38}$    &  $3.5{\pm}1.0{\times}10^{38}$ (0.45${\rm L}_{\rm Edd}$) &  $2.659^{+0.004}_{-0.01}$ & $-$ & $0.6$\\
         4 &   $1.3{\pm}0.2{\times}10^{38}$ & $3.4{\pm}0.4{\times}10^{37}$ & $1.6{\pm}0.2{\times}10^{38}$ (0.20${\rm L}_{\rm Edd}$) & $2.39{\pm}0.06$ &  $-$    & $3.8$   \\
         5 &   $1.1{\pm}0.2{\times}10^{38}$ & $2.6{\pm}0.5{\times}10^{37}$ & $1.4{\pm}0.3{\times}10^{38}$ (0.18${\rm L}_{\rm Edd}$) &  $2.65{\pm}0.02$ & $-$       & $4.0$ \\
           &  &    &   &        &   &  \\
 \enddata
 \end{deluxetable}

\begin{deluxetable}{cccccccc}
\tabletypesize{\scriptsize}
\rotate
 \tablecaption{Best-fit parameters of the joint ISGRI and SPI spectra for epochs 1, 2 and 3 by fitting data with thermal-Comptonization model and thermal-Comptonization ({\tt compps} plus a
powerlaw. Comptonization temperature ($kT$) and optical depth (${\tau}$) are free parameters; ${\Gamma}$ is the powerlaw photon-index. In COMPPS model the black body temperature
of the seed photons was frozen to the value obtained by EQPAIR fits of XMM/EPIC-pn spectra in all epochs (except of epoch 2, which was frozen to 0.77\,keV). Reflection component
(${\Omega}/2{\pi}$) is only present in COMPPS.
\label{fit_compps_param}}
 \tablewidth{0pt}
 \tablehead{\colhead{${\rm Epoch}$} & \colhead{${\rm Model}$}    & \colhead{$kT_{e}$} & \colhead{${\tau}$} & \colhead{$[{\Omega}/2{\pi}]$} & \colhead{$[{\Gamma}]$}  & \colhead{${\chi}_{\nu}^{2}$(dof)} & \colhead{F-test p}
 }
 \startdata
           &       &  [keV]    &               &                          &                &         &        \\
           &       &           &               &                          &                &         &        \\
\hline
           &       &           &               &                          &                &         &        \\
 1         &  COMPPS    & $43.6{\pm}1.4$  &  $2.89{\pm}0.11$  &   $0.92{\pm}0.08$ &   $-$             &  1.02 (54)       &  $-$      \\
 1         &  COMPPS+PO & $40.8{\pm}0.9$  &  $2.95{\pm}0.05$  &   $0.94{\pm}0.06$ &  $1.1{\pm}0.6$ &  0.92 (52)       & 6.837E-02       \\
 2         &  COMPPS    & $103{\pm}45$    &  $0.48^{+0.8}_{-0.05}$ &  0           &   $-$             &  1.10 (26)       &  $-$      \\
 2         &  COMPPS+PO & $67^{+10}_{-46}$ &  $0.41{\pm}0.36$  &  0               & $1.99^{+1.0}_{-0.07}$ &  1.16 (24)       &  $-$      \\
 3         &  COMPPS    & $135{\pm}15$ &  $0.55{\pm}0.32$  &  $0.37{\pm}0.25$  &   $-$             &  1.04 (26)       &  $-$      \\
 3         &  COMPPS+PO & $31{\pm}8$   &     $2.95{\pm}0.05$  &      $0.37{\pm}0.21$     &  $1.75{\pm}0.8$   &  1.06 (24)       &    $-$      \\
           &       &           &               &                          &                &         &        \\
 \enddata
 \end{deluxetable}

\begin{deluxetable}{ccccccccc}
\tabletypesize{\scriptsize}
\rotate
 \tablecaption{Best-fit parameters of the joint XMM/EPIC-pn, JEM-X, ISGRI and SPI spectra for the 5 epochs. Fits have been performed simultaneously with EQPAIR combined with LAOR. 
The inclination angle of the disk was frozen to the value $i=50^{\circ}$ for the first epoch and $i=30^{\circ}$ for the remainder ones. See text for the parameters description.
\label{fit_eqpair_param}}
 \tablewidth{0pt}
 \tablehead{\colhead{${\rm Epoch}$} &  \colhead{$kT_{pn}$} & \colhead{${\ell}_{h}/{\ell}_{s}$} & \colhead{${\ell}_{nth}/{\ell}_{h}$} & \colhead{${\tau}_{p}$} & \colhead{$kT_{e}$} & \colhead{${\Gamma}_{\rm inj}$} & \colhead{$[{\Omega}/2{\pi}]$} & \colhead{${\chi}_{\nu}^{2}$(dof)}
 }
 \startdata
           &   [eV]    &               &                          &                &  [keV]  &        &      &  \\
           &   &       &            &      &    & & \\
\hline
           &   &       &            &      &    & & \\
 1         & 300 (f)  & $3.9^{+0.6}_{-0.2}$ &  $0.40_{-0.03}^{+0.15}$ & $2.39_{-0.18}^{+0.15}$ & $27.5{\pm}1.2$  & $1.80^{+0.07}_{-0.05}$  & $0.38^{+0.06}_{-0.04}$ & $1.0$(82) \\
 2         & $897_{-1.3}^{+1.4}$  & $0.05_{-0.01}^{+0.003}$ & $0.90{\pm}0.10$ & $<0.02$ & $69{\pm}4$ & $2.87^{+0.05}_{-0.06}$   & $0.40_{-0.04}^{+0.3}$   & $1.7$(142) \\
 3         & $704{\pm}2$  & $0.28^{+0.03}_{-0.01}$ & $0.84{\pm}0.03$ & $1.41_{-0.06}^{+0.03}$ &   $10.8{\pm}0.3$ & $3.42_{-0.04}^{+0.03}$   & 1(f) & $1.5$(138) \\
 4         & 500(f)  & $0.24^{+0.02}_{-0.005}$ & $0.49^{+0.02}_{-0.01}$ & $2.5{\pm}0.5$ & $4.3{\pm}0.8$  & $2.36^{+0.04}_{-0.08}$ & 1(f) & 1.5(44) \\
 5         & $563{\pm}2$  & $0.13^{+0.012}_{-0.018}$ & $0.38{\pm}0.02$  & $0.89^{+0.04}_{-0.05}$ &  $10.5{\pm}0.7$ & $2.64_{-0.09}^{+0.08}$   & $0.72^{+0.16}_{-0.10}$ & $1.4$(99) \\
           &   &       &            &      &    & & \\
 \enddata
 \end{deluxetable}




\setcounter{figure}{0}


\begin{figure}
\centering
\includegraphics[angle=0,width=1.0\linewidth]{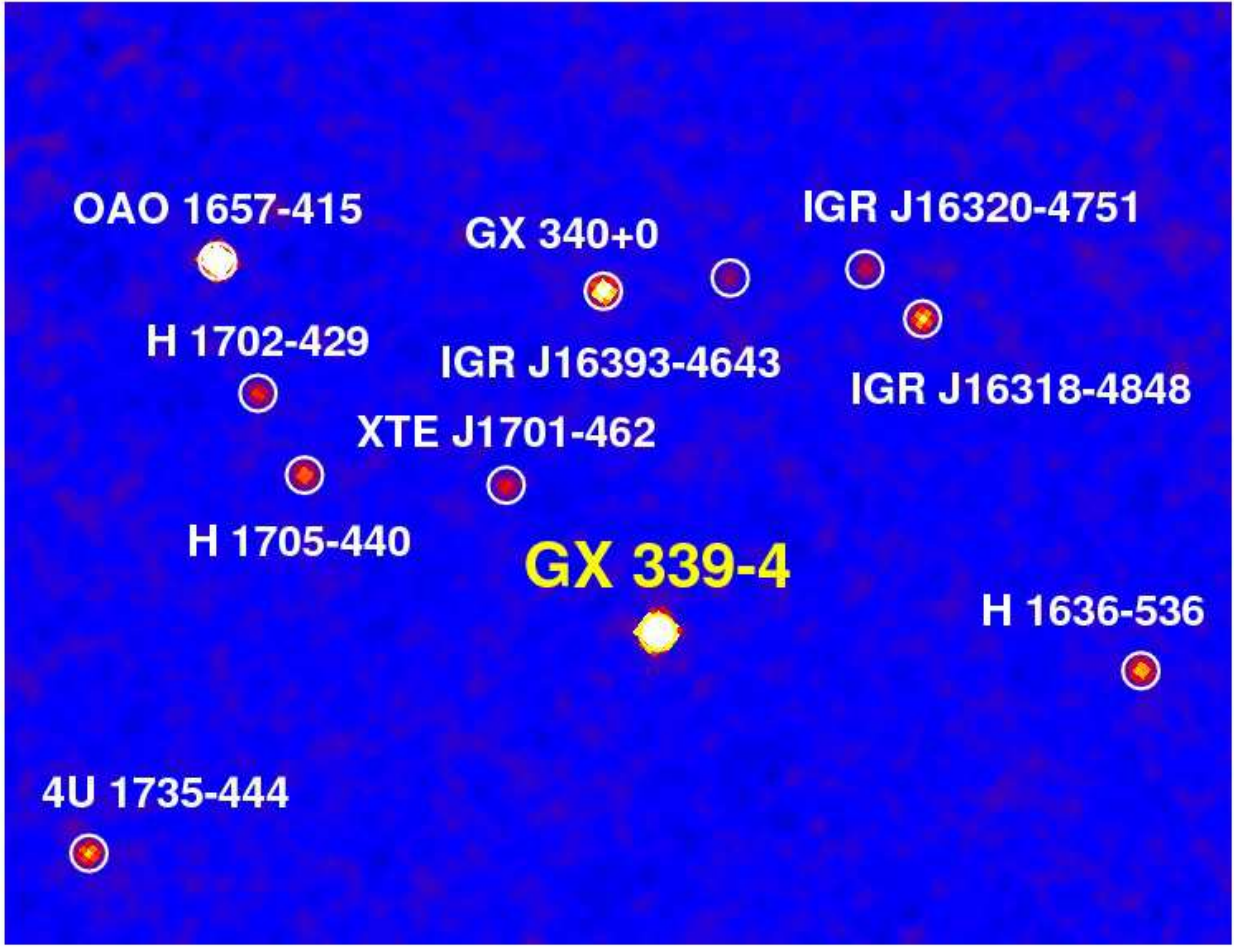}
\caption{Mosaic significance image (obtained in epoch 1) of the GX~339--4 region as seen with ISGRI in the $20-40$\,keV energy range, galactic coordinates from right to left. Besides GX~339--4 (detection isgnificance of $1\,070{\sigma}$), several other high energy sources are visible. \label{mosaic}}
\end{figure}

\begin{figure*}
\centering
\includegraphics[angle=0,width=1.0\linewidth]{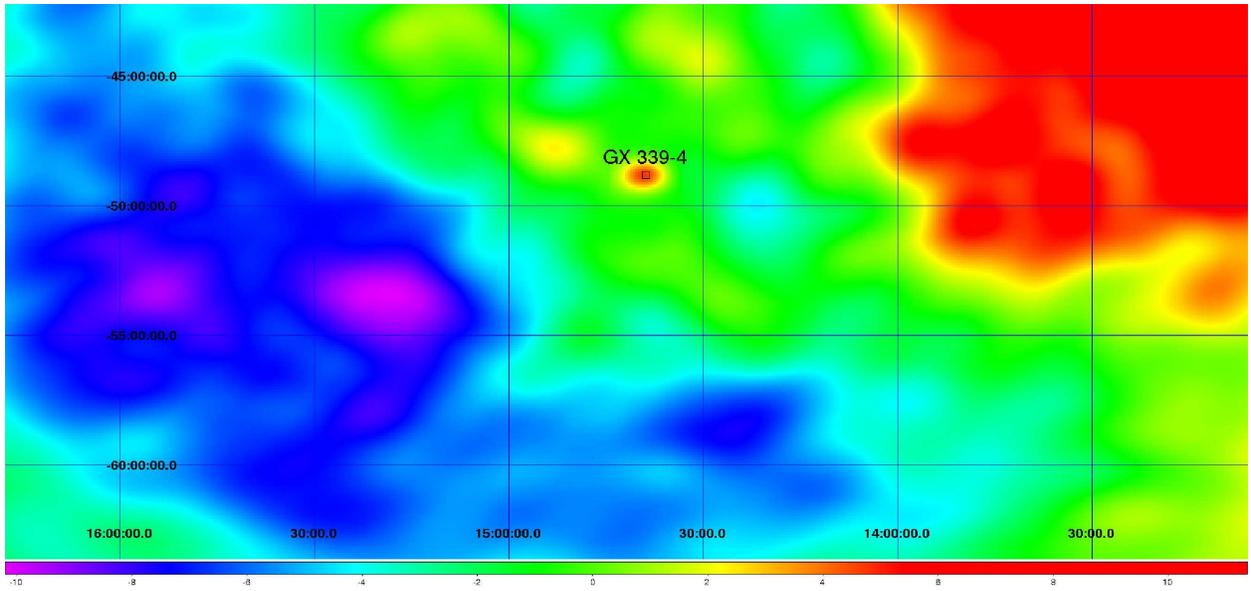}
\caption{Significance map obtained with SPIROS in the 200--300\,keV energy range during epoch 1 observations; the color scale goes from -5\,${\sigma}$ to 5\,${\sigma}$; the grid spacing is of
1\,$^{\circ}$ and 5\,$^{\circ}$ in right-ascension and declination, respectively. GX~339--4 is the only source detected in the 200--300\,keV energy range, with a flux of $0.01951{\pm}0.004030$\,ph/cm$^{2}/$sec
(4.8\,${\sigma}$).
}
\label{spi_Hchan}
\end{figure*}

\begin{figure}
\centering
\includegraphics[width=0.4\linewidth]{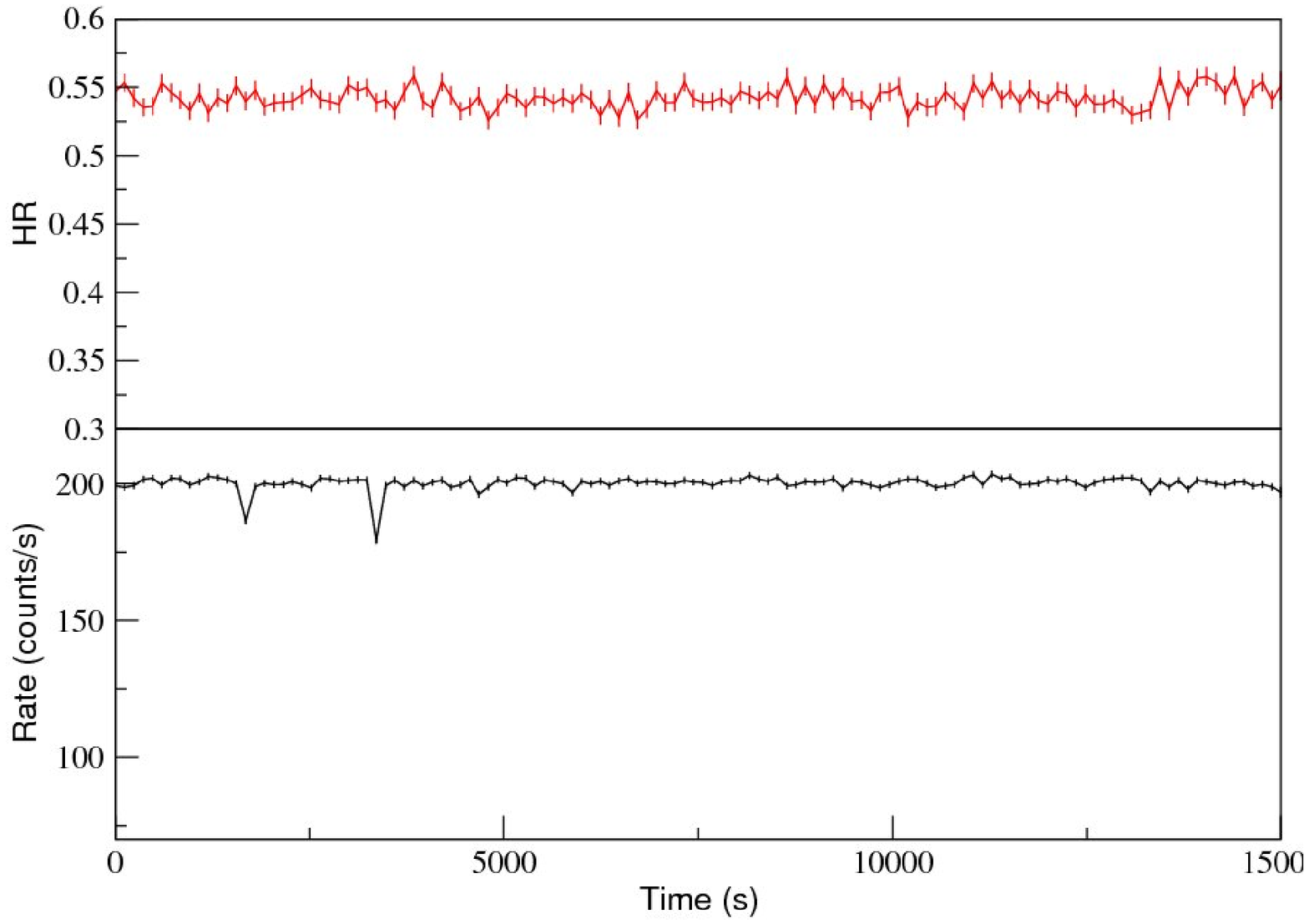}
\includegraphics[width=0.4\linewidth]{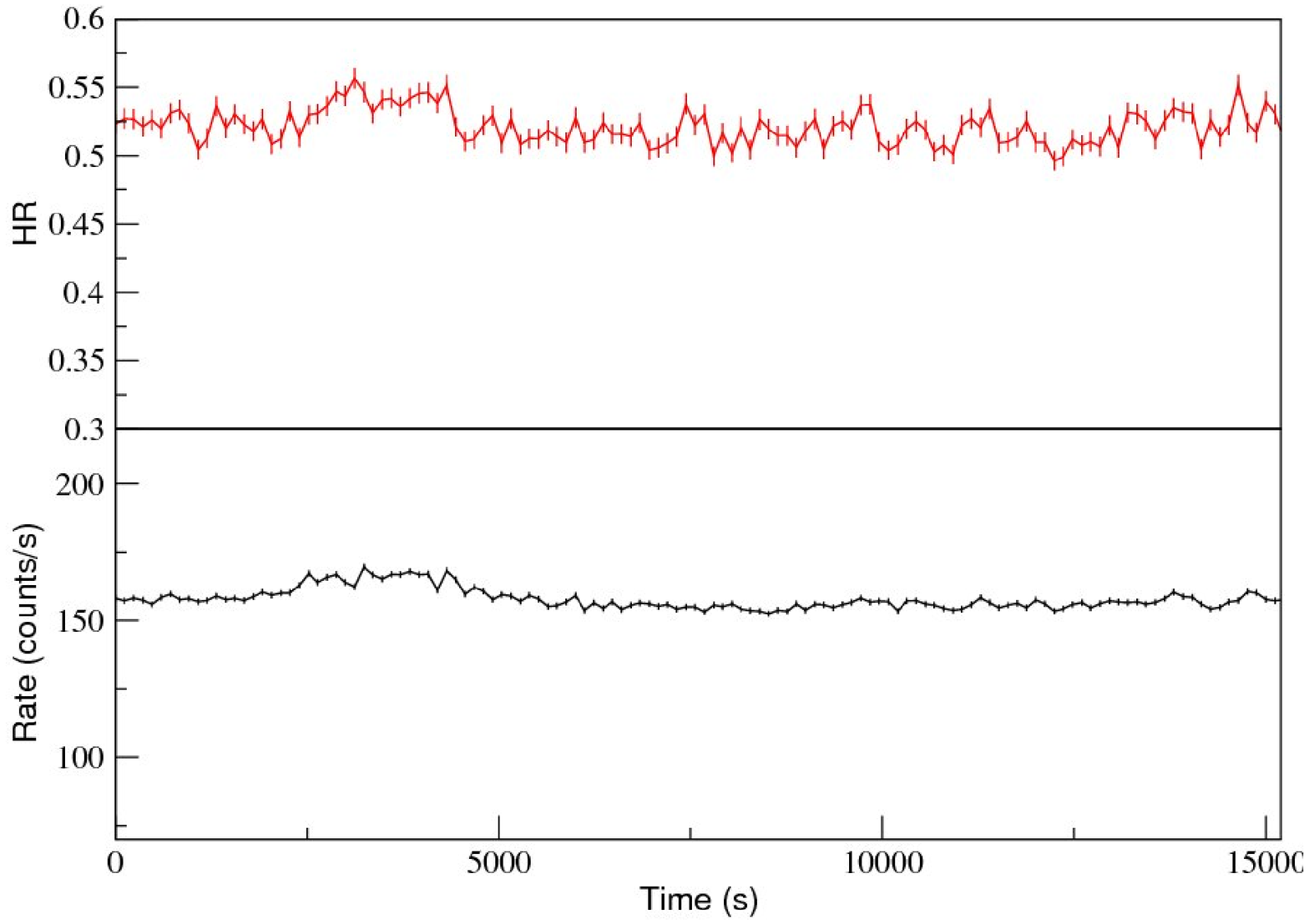}
\includegraphics[width=0.4\linewidth]{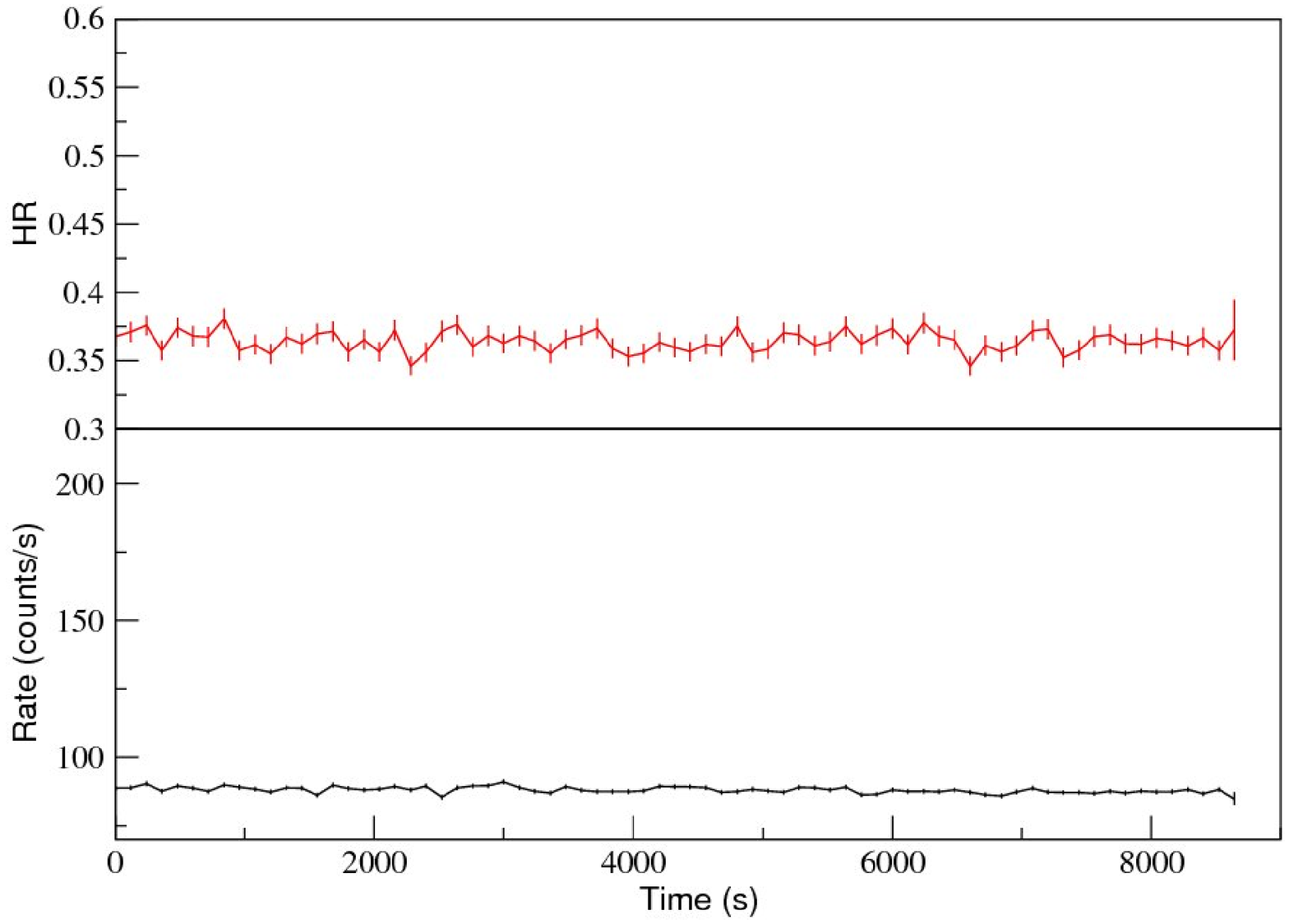}
\caption{{\it Lower panels:} 0.6-10\,keV EPIC pn light curve of GX 339--4 for epochs 2 (upper-left panel), 3 (upper-right panel) and 5 
(bottom panel).{\it Upper panels:} hardness ratio (counts in the 2--10\,keV band divided by those between 0.6--2\,keV) for epochs 2, 3 and 5. The binning is
120\,s for all panels.}
\label{lcurves_xmm}
\end{figure}

\begin{figure}
\centering
\includegraphics[width=1.0\linewidth]{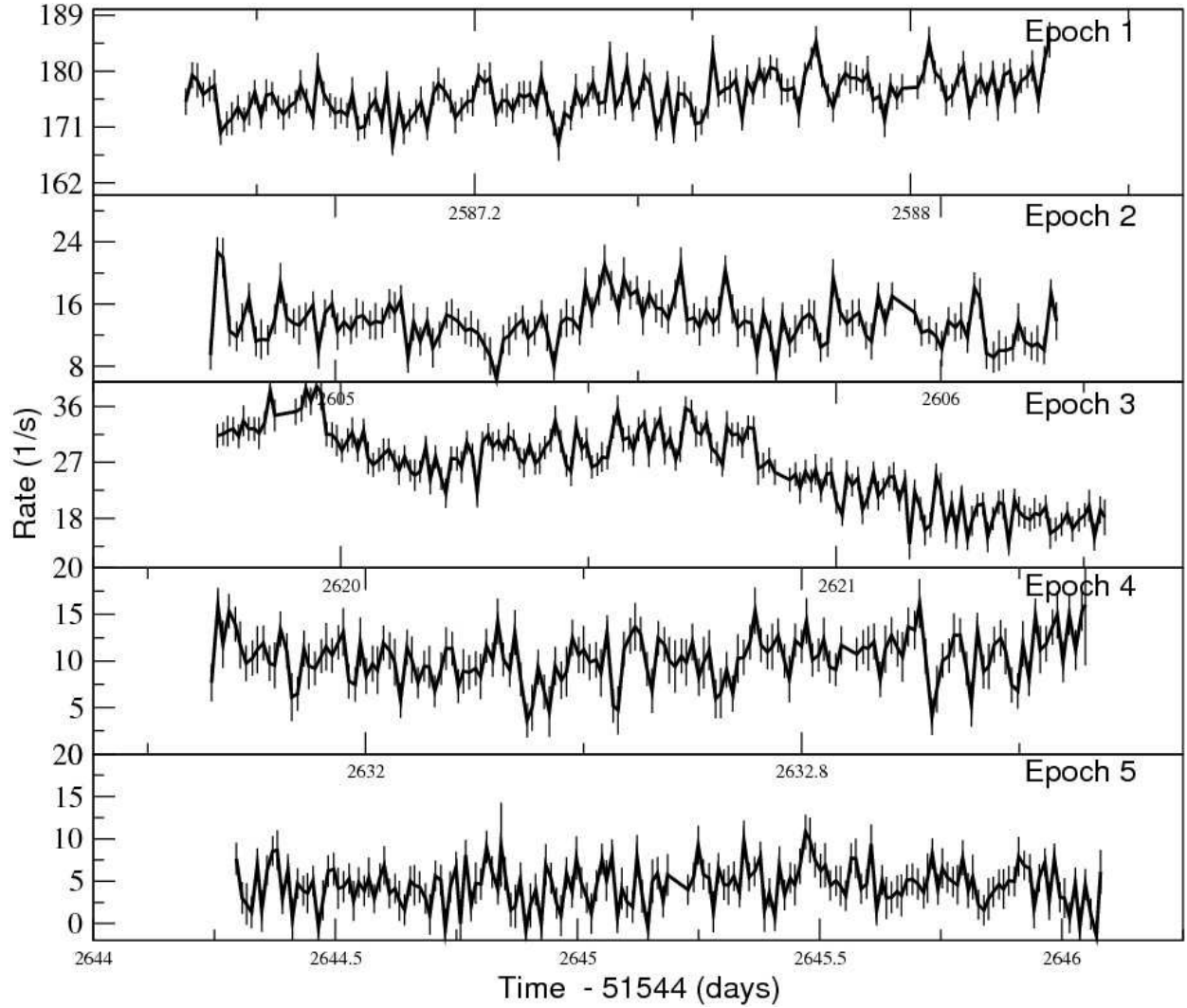}
\caption{ISGRI/INTEGRAL (20-200\,keV) light curve during all the observations (from Epoch 1 to 5). The binning is 1\,000 s for all panels.
}
\label{lcurves_integral}
\end{figure}

\begin{figure}
\centering
\includegraphics[width=1.0\linewidth]{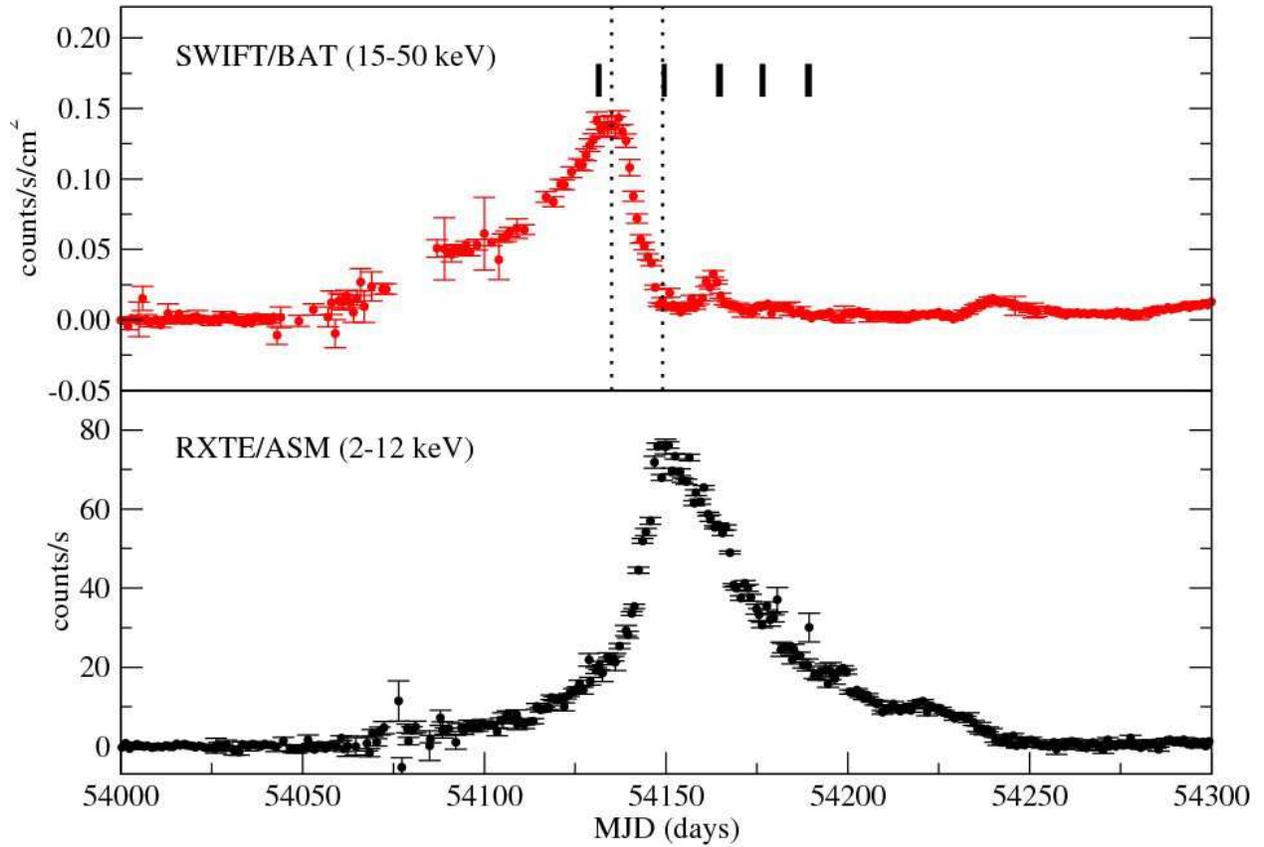}
\caption{SWIFT/BAT and RXTE/ASM daily light curves of GX~339--4 during the overall outburst in 2007 (red and black dotted lines, respectively), illustrating
the spectral evolution between the different states. Intervals of time in which the INTEGRAL observations
were undertaken (solid black lines) and the period of time when the radio ejection events were detected (within both dotted black vertical lines) are also shown.}
\label{lcurves_swift}
\end{figure}

\begin{figure}
\centering
\includegraphics[angle=0,width=0.6\linewidth]{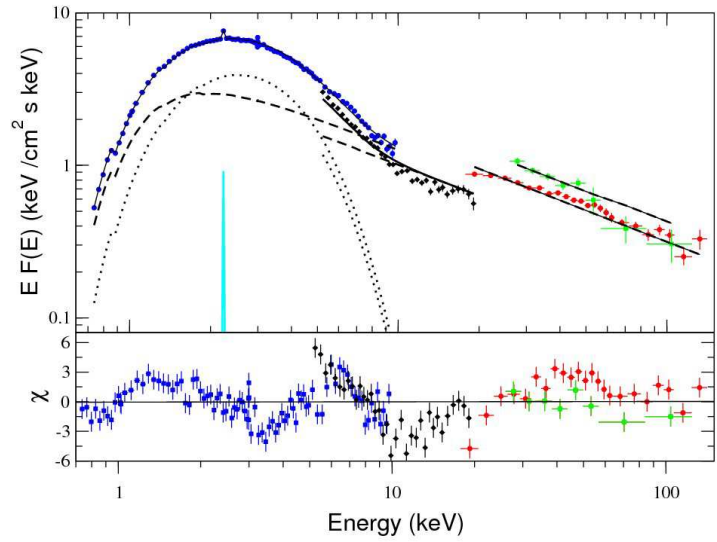}
\caption{The plot above shows the residuals obtained when the {\it INTEGRAL} and {\it XMM-Newton} data of epoch 3 is fit with a phenomenological disk
plus power-law model. In the residuals a relativistically broadened ${\rm Fe}$\,${\rm K}_{\alpha}$ line is present.}
\label{skewed_Fe}
\end{figure}

\begin{figure*}
\centering
\includegraphics[angle=0,width=0.4\linewidth]{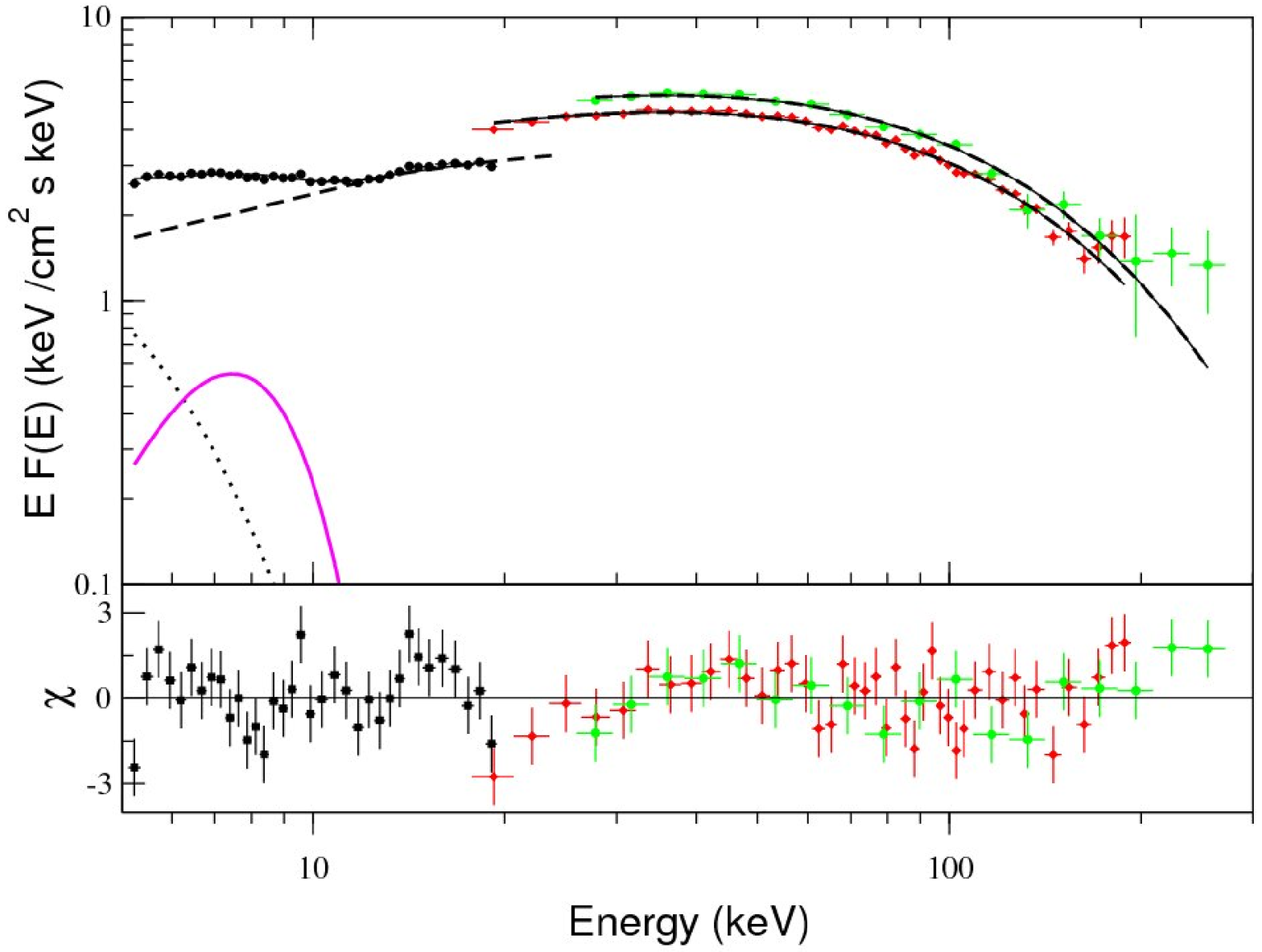}
\includegraphics[angle=0,width=0.4\linewidth]{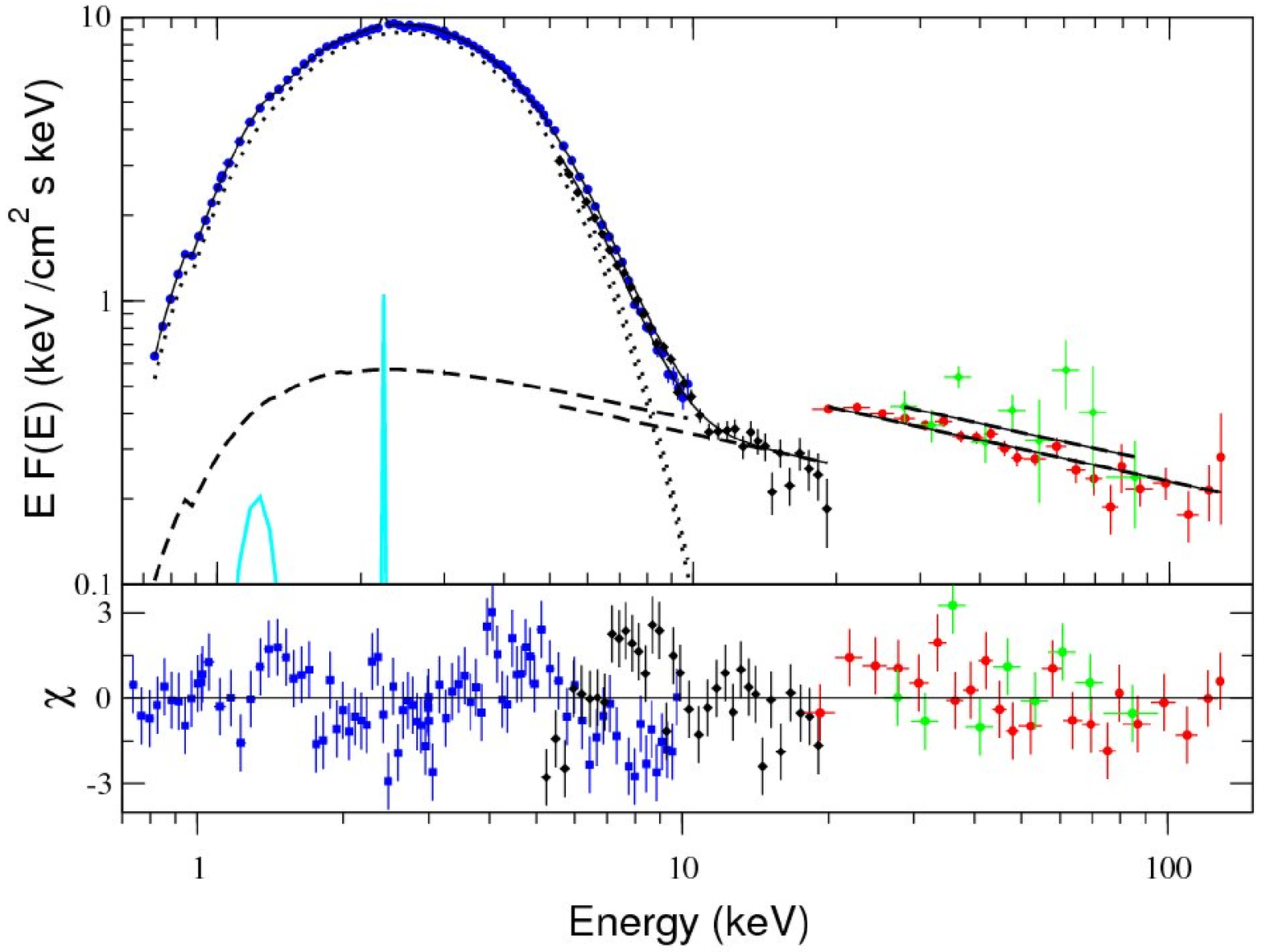}
\includegraphics[angle=0,width=0.4\linewidth]{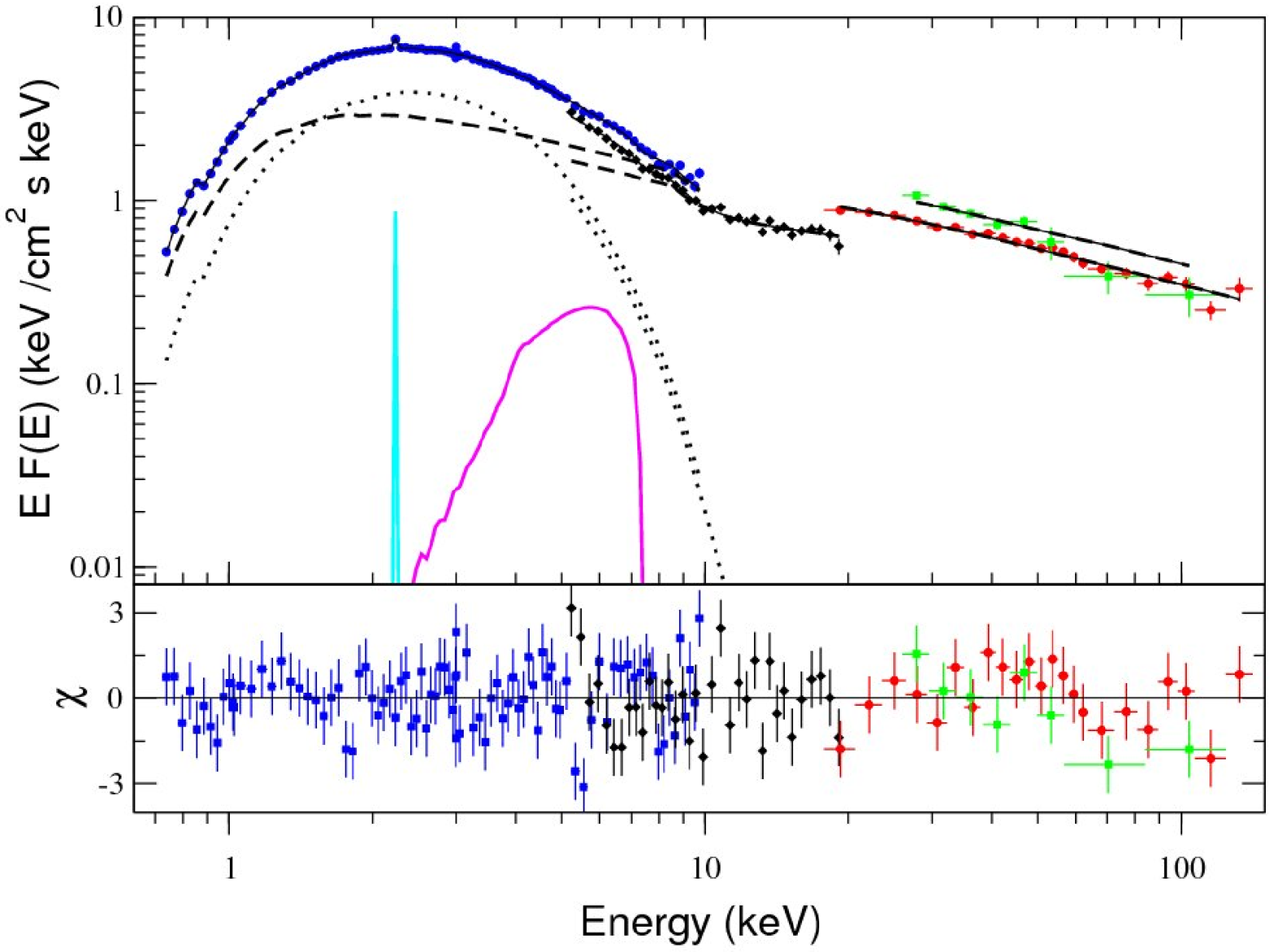}
\caption{Unfolded INTEGRAL and XMM/EPIC-pn spectra (blue, black, red and green for the XMM/EPIC-pn, JEM-X, ISGRI and SPI data, respectively) from epochs 1 to 3 
(upper-left to lower-right) fit with descriptive phenomenological models. The continuum line shows the 
total model (see the text Section \ref{spec_phenom} and Table \ref{param_spec} for
details), the dotted and dashed lines show the accretion disk and the powerlaw components, respectively, and the magenta line represents
relativistically broadened ${\rm Fe}$${\rm K}_{\alpha}$ emission from a disk. The cyan colored gaussian lines mean instrumental
XMM/EPIC-pn effects.} 
\label{fit_spec1}
\end{figure*}

\begin{figure}
\centering
\includegraphics[angle=0,width=0.4\linewidth]{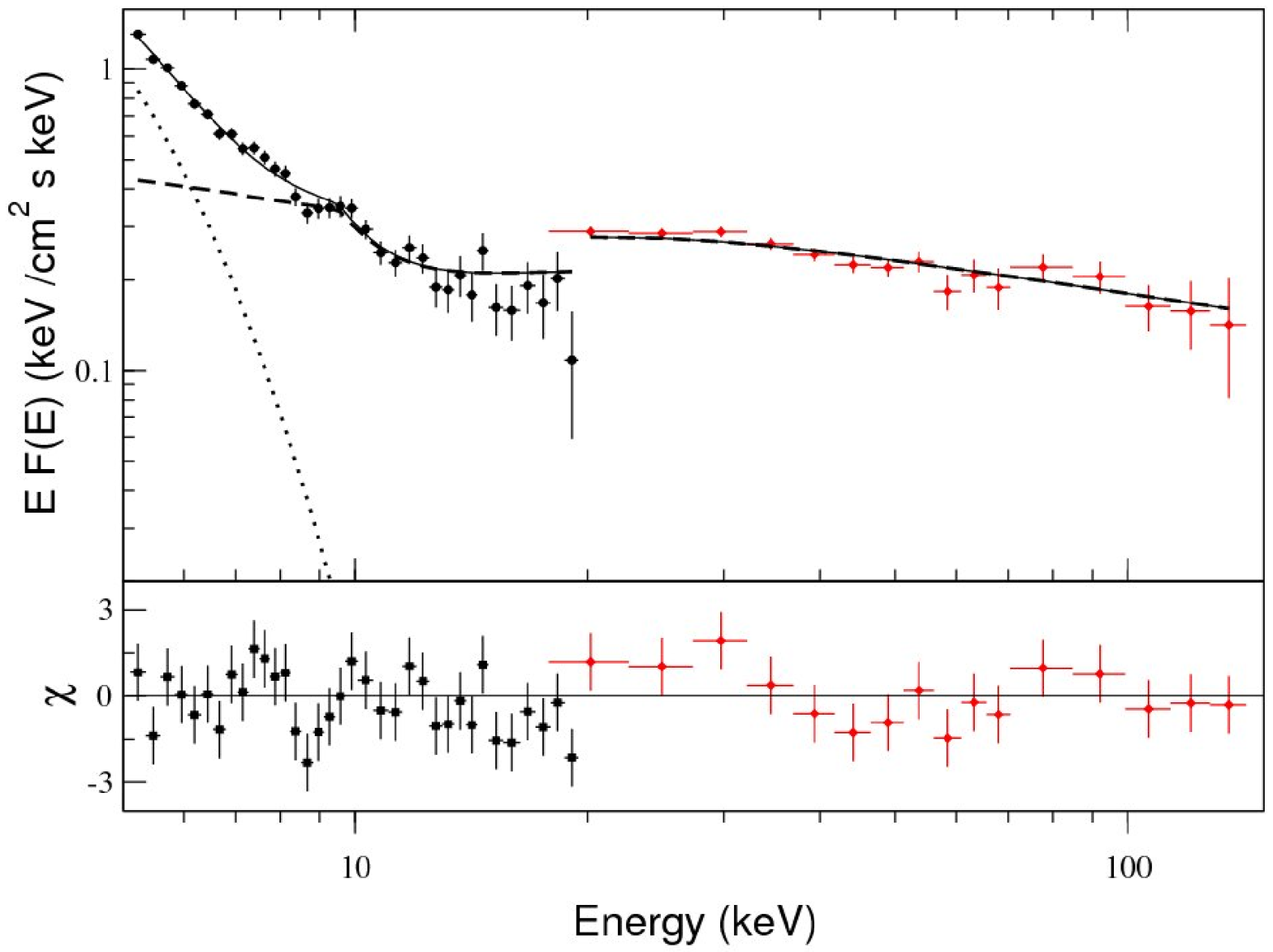}
\includegraphics[angle=0,width=0.4\linewidth]{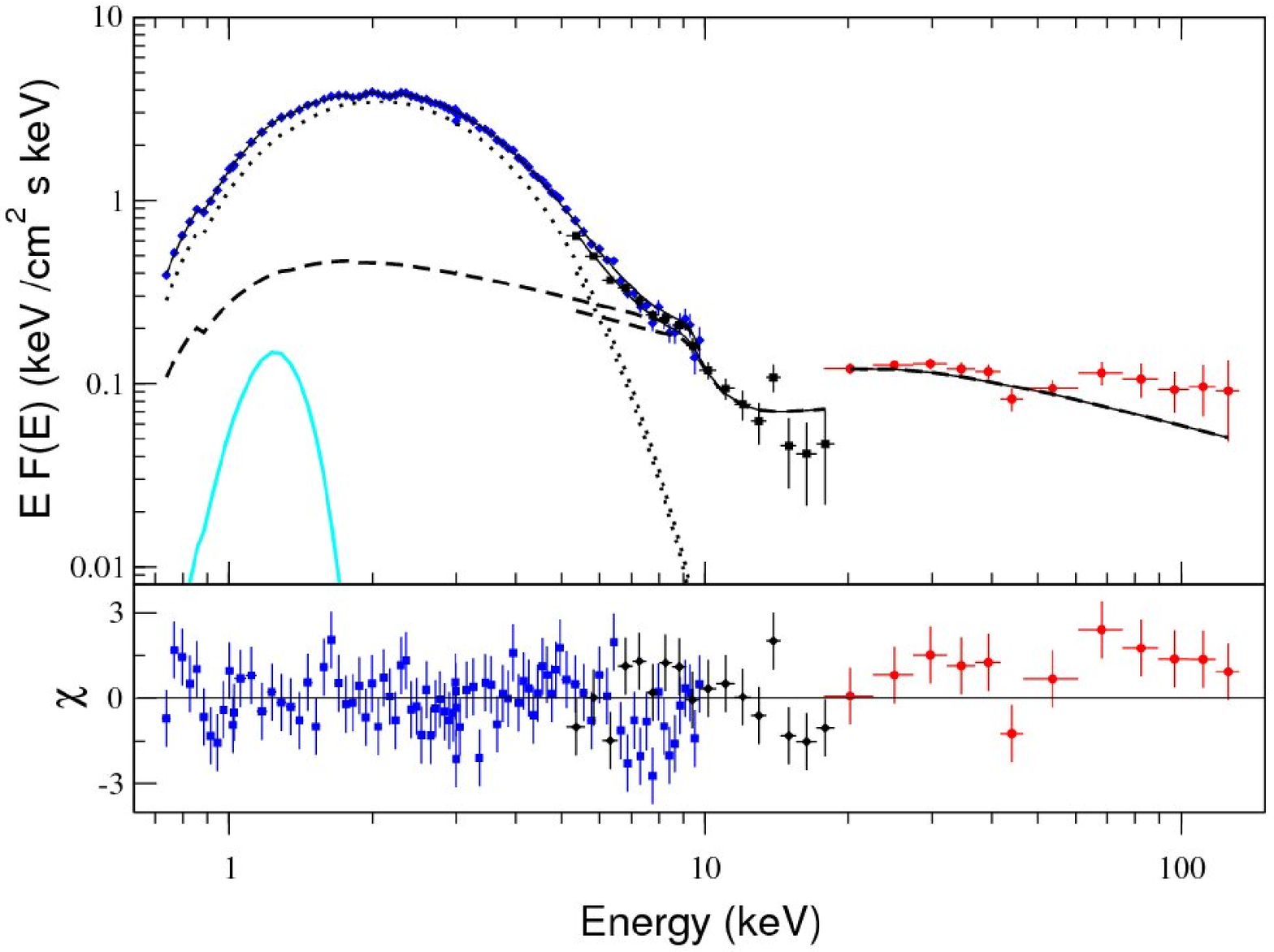}
\caption{For epochs 4 and 5 spectral evolution occurred in the form of a gradual softening. See Section \ref{spec_phenom} for more details.} 
\label{fit_spec2}
\end{figure}

\clearpage

\begin{figure}
\centering
\includegraphics[angle=0,width=1.0\linewidth]{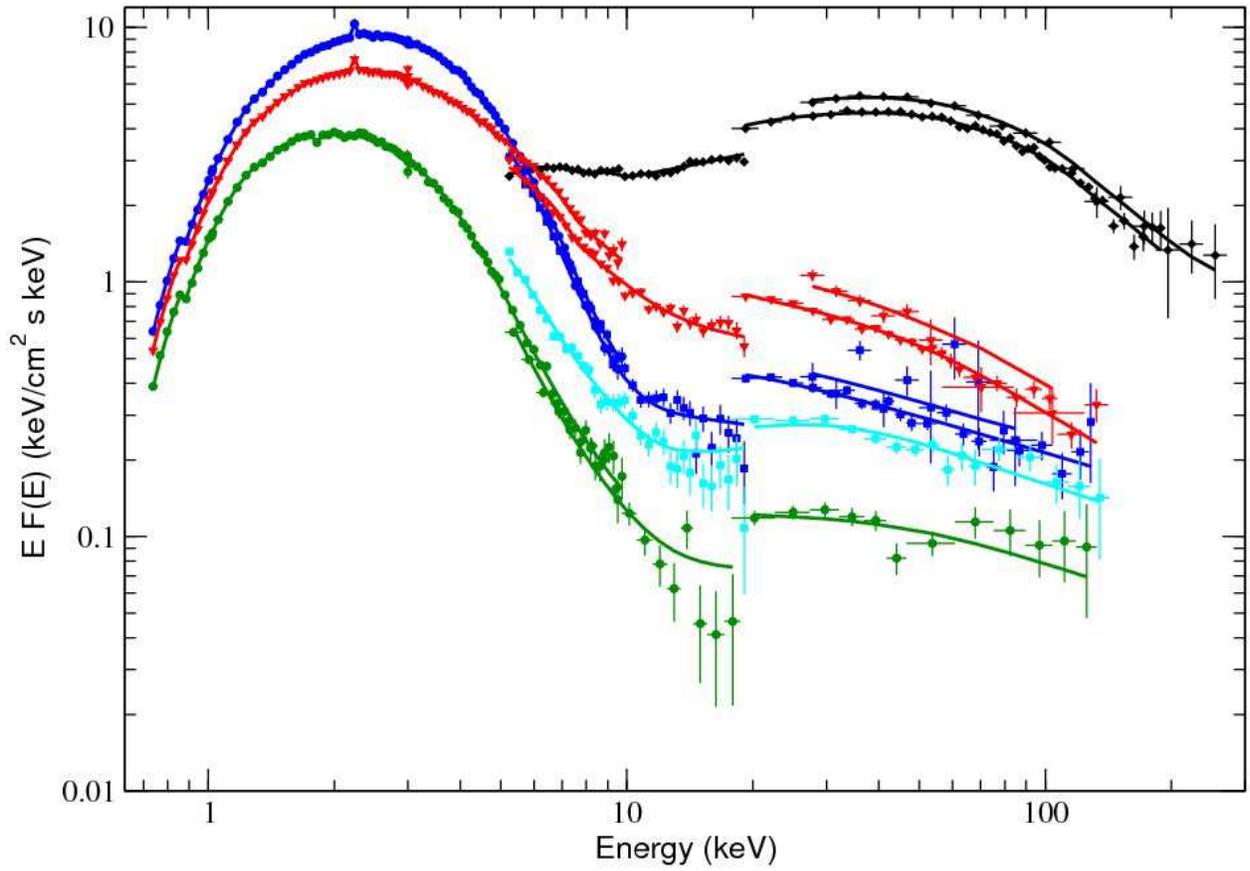}
\caption{Unfolded spectra from epoch 1 to 5 (black, blue, red, cyan and green, respectively).
}
\label{spec_state}
\end{figure}

\begin{figure}
\centering
\includegraphics[angle=270,width=0.4\linewidth]{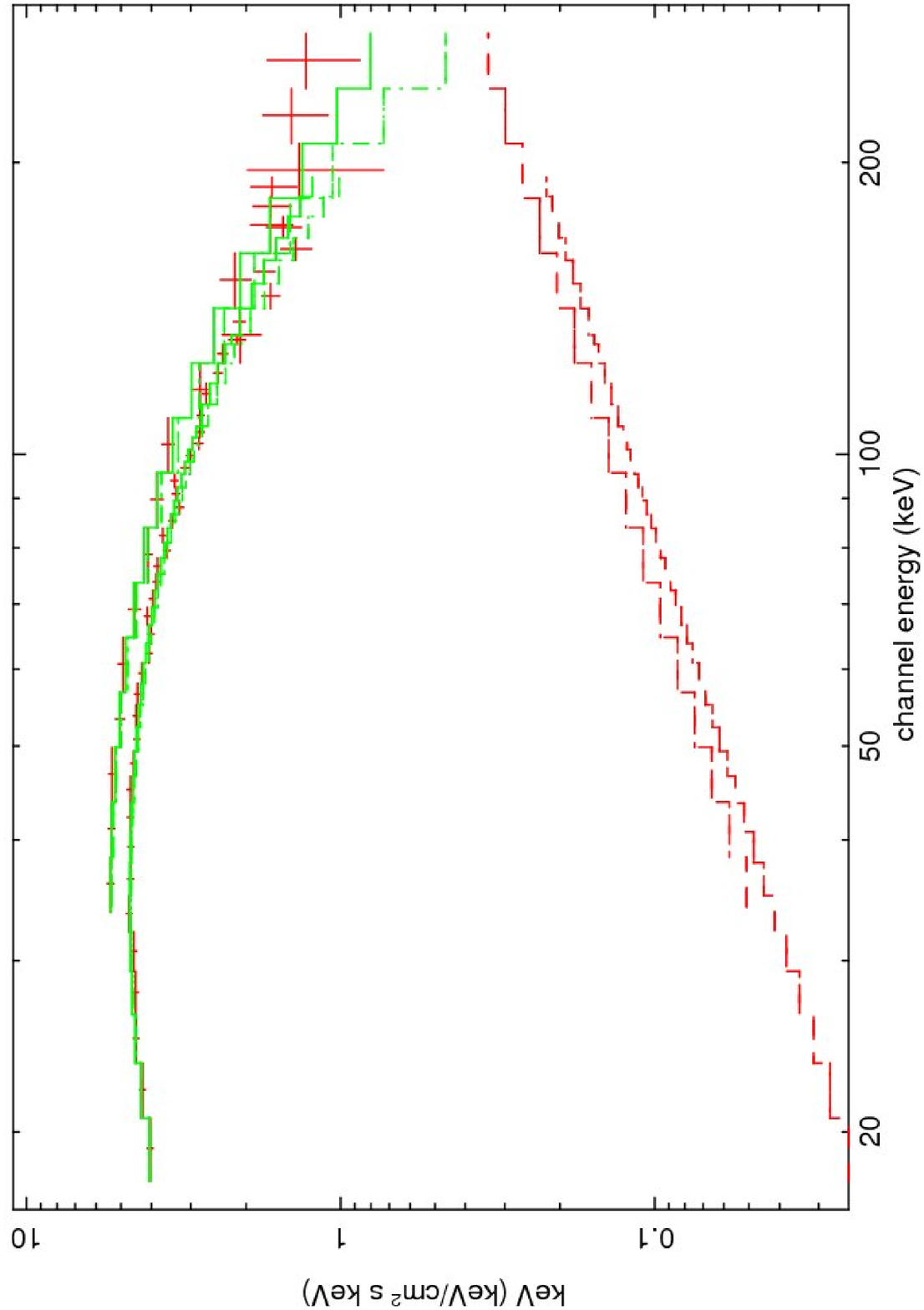}
\includegraphics[angle=270,width=0.4\linewidth]{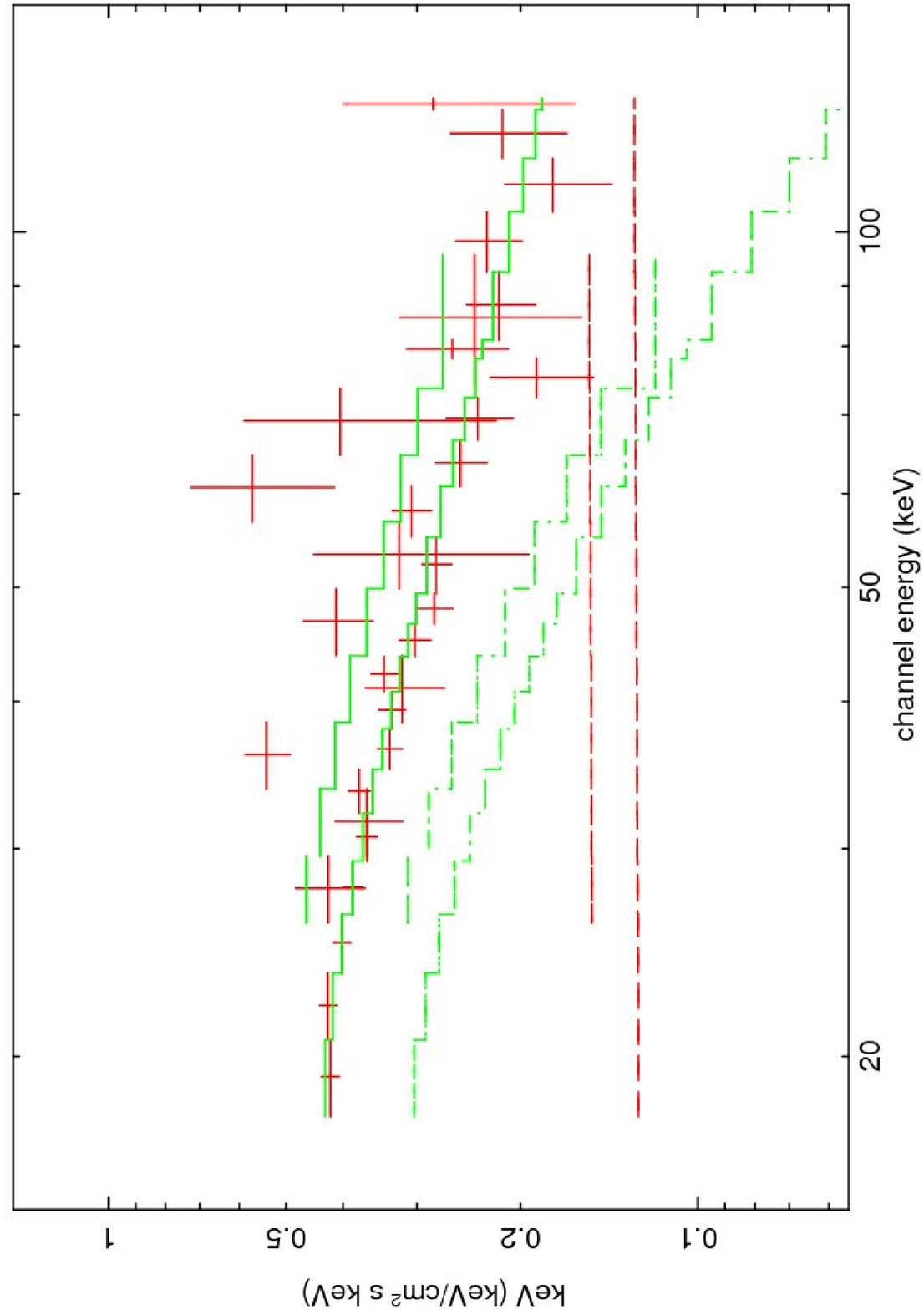}
\includegraphics[angle=270,width=0.4\linewidth]{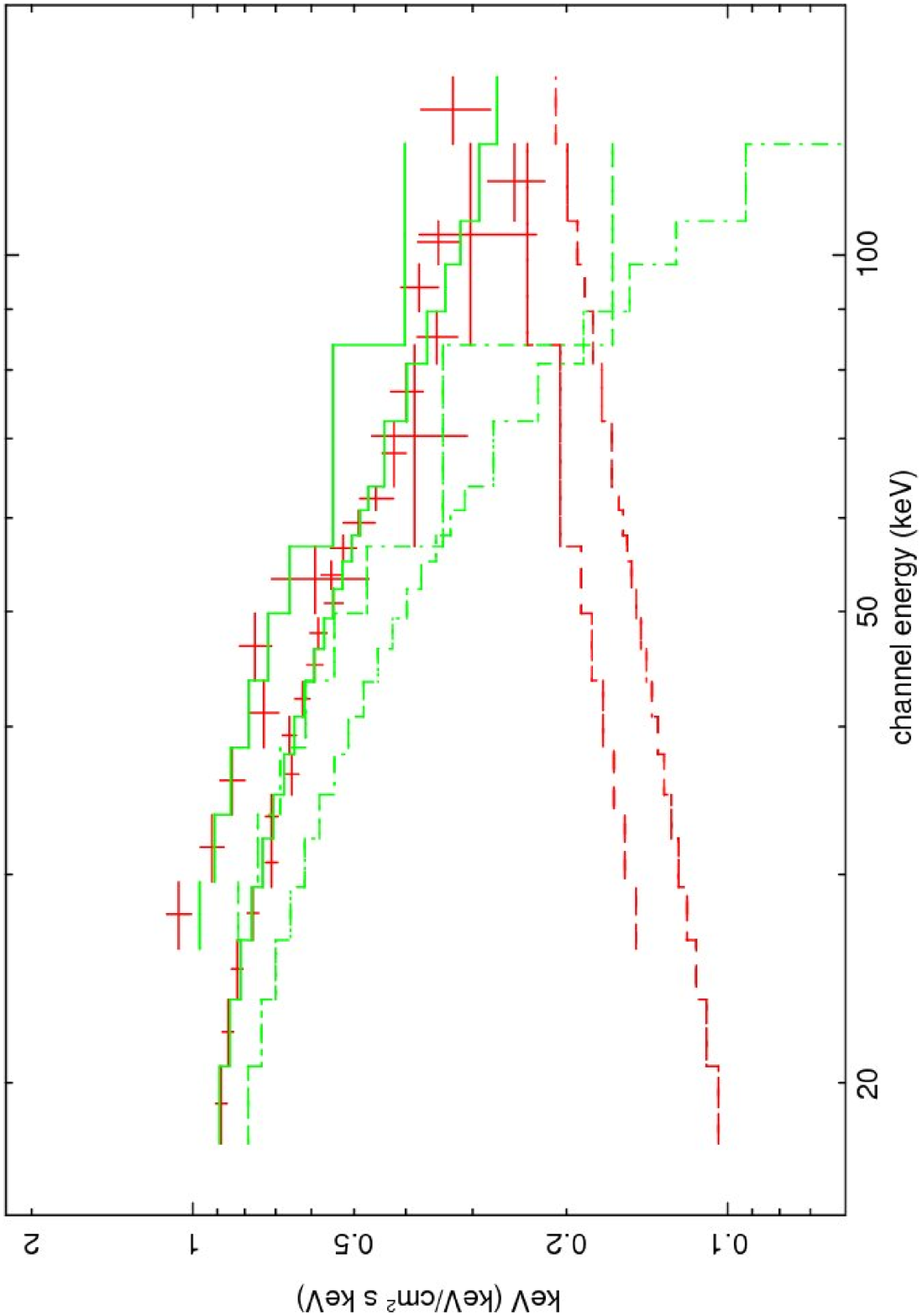}
\caption{IBIS and SPI unfolded energy spectra for epochs 1, 2 and 3 (left to right) fit with a Comptonization model (COMPPS) plus a powerlaw.}
\label{fit_compps}
\end{figure}

\end{document}